\documentclass[useAMS,usenatbib]{mn2e}

\usepackage{placeins}

\usepackage{graphicx} 
\usepackage{multirow}
\usepackage{array}

\usepackage{amsmath,amssymb}

\usepackage[T1]{fontenc}
\usepackage[latin9]{inputenc}
\usepackage{float}
\usepackage{graphicx}
\usepackage{esint}

\usepackage{auto-pst-pdf}

\makeatletter
\DeclareFontEncoding{LGR}{}{}
\DeclareTextSymbol{\~}{LGR}{126}

\begin{document}

\title[Relative velocity signatures]{On the signature of the baryon-dark matter relative velocity in the
two and three-point galaxy correlation functions}

\author{\makeatauthor}

\author[Slepian and Eisenstein]{Zachary Slepian\thanks{E-mail: zslepian@cfa.harvard.edu} and Daniel J. Eisenstein\thanks{E-mail: deisenstein@cfa.harvard.edu}\\
Harvard-Smithsonian Center for Astrophysics, Cambridge, MA 02138\\
}
\maketitle

\begin{abstract}
 We develop a configuration-space
picture of the relative velocity between baryons and dark matter that clearly explains how it can shift the BAO scale in the galaxy-galaxy correlation
function. The shift occurs because the relative velocity is non-zero only within the sound horizon and thus adds to the correlation function asymmetrically about the BAO peak.  We further show that in configuration space the relative velocity has a localized, distinctive signature in the three-point galaxy correlation function (3PCF). In particular, we find that a multipole decomposition is a favorable way to isolate the relative velocity in the 3PCF, and that there is a strong signature in the $l=1$ multipole for triangles with 2 sides around the BAO scale. Finally, we investigate a further compression of the 3PCF to a function of only one triangle side that preserves the localized nature of the relative velocity signature while also nicely separating linear from non-linear bias. We expect that this scheme will substantially lessen the computational burden of finding the relative velocity in the 3PCF. The relative velocity's 3PCF signature can be used to correct the shift induced in the galaxy-galaxy  correlation function so that no systematic error due to this effect is introduced into the BAO as used for precision cosmology.
\end{abstract}

\section{Introduction}
The baryon acoustic oscillation (BAO) method uses the imprint of sound waves in the early Universe on the clustering of galaxies today as a sensitive probe of the Universe's expansion history (see Weinberg et al. 2013 for a recent review).  This in turn constrains the dark energy equation of state, which offers insight into dark energy's fundamental nature (Albrecht et al. 2006 for a review; Copeland et al. 2006; Li et al. 2011 for model compendia; for recent work on specific models, see e.g. Dutta \& Scherrer 2008; Chiba 2009; Chiba et al. 2009; Gott \& Slepian 2011; De Boni et al. 2011; Chiba et al. 2013; Slepian et al. 2014).  The BAO method's accuracy depends on precisely modeling how the sound waves frozen in at high redshift imprint on galaxy clustering today, and hence how baryons and DM combine to form these galaxies.  

A potentially important effect on early generations of galaxies is the supersonic relative velocity between baryons and DM at decoupling
($z\sim1020)$, recently presented by Tseliakhovich \& Hirata (2010). The relative velocity is sourced by the difference in the behavior
of baryons and dark matter before decoupling. Prior to decoupling, the baryons and photons form a tightly coupled fluid, locked together by Thomson scattering (linking electrons to photons) and the Coulomb force (linking protons to electrons).  This fluid undergoes acoustic oscillations, or sound waves, that propagate to roughly 150 Mpc comoving before halting as electron-photon scattering drops precipitously and decoupling occurs (Peebles \& Yu 1970; Sunyaev \& Zel'dovich 1970; Bond \& Efstathiou 1984, 1987; Holtzmann 1989; Hu \& Sugiyama 1996; Eisenstein \& Hu 1998). The scale at which these waves halt is termed the sound horizon.

Given an isolated overdense region, baryons nearer to it than the sound horizon are kept in rough hydrostatic equilibrium by the radiation pressure and so do not infall.  In contrast, baryons more distant than the sound horizon fall towards the overdensity. Meanwhile, DM on {\it all} scales infalls gravitationally.
Consequently, the baryons and DM differ in behavior below the sound horizon, resulting in a relative velocity at decoupling on these scales.
 
It is believed that the relative velocity can modulate the formation of the first galaxies in the Universe on scales similar to the sound horizon (we describe this more below).  Since these galaxies are the progenitors of those we observe today, galaxy clustering today may retain a memory of this effect.  Yoo et al. (2011) analyze how such a memory might cause a shift in the galaxy-galaxy correlation function by which the sound horizon scale today is measured, an idea also hinted at in Tseliakhovich \& Hirata 2010 and Dalal et al. 2010. Given the high precision of current
and impending BAO surveys such as BOSS, even a modest, order $1\%$
systematic source of error could significantly bias the inferred cosmological
parameters. Therefore it is essential to understand how the relative
velocity can induce this shift and how, if the shift is indeed present,
it can be corrected.  While Yoo et al. (2011) state that the correlation function can shift, their analysis presents results in Fourier space, showing the power spectrum and, importantly, finding that the bispectrum can be used to remove the relative velocity effect from the power spectrum.

Here, we focus on configuration space, for several reasons.  First, complicated behavior in the power spectrum and bispectrum often has
a simple interpretation in configuration space (Bashinsky \& Bertschinger
2001, 2002). Our work shows that
the relative velocity is indeed simple in configuration space: it
is non-zero only within the sound horizon. Our work therefore makes it clear that any effect on the
correlation function is primarily on sub-horizon scales. It is adding
or subtracting from the correlation function only inward of the BAO peak
that shifts the peak in or out in scale. 

Our configuration space
approach also offers the new result that, for extracting the relative
velocity from the three-point galaxy correlation function (3PCF), Legendre polynomials are an excellent angular basis (Szapudi 2004 and Pan \& Szapudi 2005 first suggested such a basis for general measurements of the bispectrum). Because the relative velocity has compact
support in configuration space, we can additionally integrate
over one side-length of the triangles entering the 3PCF for lengths where the relative
velocity has support. This produces a novel compression scheme that
improves the chances for detecting the relative velocity in the 3PCF while easing the computational demands of such an effort.

Indeed, this compression scheme is not the only practical advantage of a configuration space approach.  The bispectrum is challenging to measure accurately
on a cut sky, because survey boundaries break the translational symmetry implicit
in a Fourier decomposition.  They also impose some miminum wavenumber
below which the Fourier representation is truncated, leading to Gibbs
phenomenon ringing in the bispectrum. In contrast, in configuration
space, the 3PCF can be measured straightforwardly and cut-sky effects
corrected by use of an estimator (see e.g. Kayo et al. 2004; Szapudi
2004; Szapudi and Szalay 1998), though at some computational cost
(McBride et al. 2011; Marin et al. 2013). Indeed, Pan and Szapudi
(2005) have already measured the monopole moment of the 3PCF in Two-degree-Field Galaxy Redshift Survey (2dFGRS),
showing the feasibility of this approach.

In the remainder of the Introduction, we give greater detail on the physical
mechanisms by which the relative velocity may affect galaxy clustering today: how does the relative velocity effect the first galaxies to form, and how might galaxies today retain a memory of these distant progenitors?  The relative velocity affects the formation of early, low-mass haloes,
but precisely how remains an open question. In their initial paper
presenting the relative velocity, Tseliakhovich \& Hirata (2010) predict
suppression of halos with $M\lesssim10^{6}\; M_{\odot}$ due to the
relative velocity.  Naoz et al. (2012) find this in simulations as well, though simulations by Richardson  et al. (2013) find only a small effect in halo number density by $z\sim20$. In
those halos that do form, the gas content is lowered (Dalal et al.
2010; Tseliakhovich et al.
2011; Fialkov et al. 2012; Naoz et al. 2013). In simulations, Maio
et al. (2011) find that star formation in low mass halos is suppressed,
though Stacy et al. (2011) argue that the later-time star formation
is not strongly affected. The minimum cooling mass for star formation
via molecular hydrogen lines also may be raised in simulations (Greif et al. 2011,
though Stacy et al.'s earlier work argues it is not). O'Leary
and McQuinn (2012) simulate structure formation to show that the relative
velocity has a substantial effect on the first mini-halos' accretion
history. Barkana (2013) points out that there may be additional dynamical
effects, such as asymmetric disruption of accreting gas filaments
and formation of supersonic wakes by halos moving in regions of high
relative velocity. Bovy \&
Dvorkin (2013) suggest that for reasons such as these, star formation
in small DM halos may be suppressed enough to resolve the over-prediction
of small halos in $\Lambda{\rm CDM}$ simulations.

It is believed that the modulation of early, low-mass halos by the
relative velocity as discussed above will affect the subsequent formation of the higher-mass
halos we observe today, perhaps through feedback channels such as altering the metal
abundance or supernovae rate (Yoo et al. 2011). Since these links
are not known in detail, it is simply assumed that the relative velocity
biases the galaxy overdensity with some amplitude $b_{v}$, to be fit from
the data. This will be our approach here as well.

Finally, numerous studies have also developed rich small-scale
consequences of the relative velocity, though that will not be our focus here.  To give only a few examples, Naoz \& Narayan
(2013) show that the relative velocity modifies the Biermann battery
picture of magnetogenesis, while Tanaka et al. (2013), Tanaka \& Li (2014), and Latif et al. (2014) consider the impact on primordial supermassive black hole formation.  Much work has also investigated the consequences
of the relative velocity for the 21 cm radiation field, e.g. Visbal
et al. (2012); McQuinn \& O'Leary (2012).  A detailed recent review of work on the relative velocity is Fialkov (2014).

This paper is structured as follows. In \S2, we lay out our approach and assumptions. In \S3, we present the structure of the relative velocity
due to a point perturbation (the Green's function), and in \S4 we
compute the shift the relative velocity induces in the correlation function. \S5 discusses
this shift and shows how it can be traced back to the compact support
of the Green's function. \S6 presents the 3PCF at one vertex of a
triangle of galaxies, and \S7 connects this with the sum over all
vertices that we observe and shows how the 3PCF may be compressed
to maximize the signal. \S8 concludes. An Appendix presents mathematical
results that we use in the paper to accelerate the numerical calculations
of \S4.

\section{Approach and assumptions}
We begin by presenting our bias model and
then outline how the spatial structure of the relative velocity (which
this model requires as an input) can be found using a Green's function
approach.

Throughout this paper, we use linear perturbation theory in configuration
space and neglect redshift-space distortions.
We model the low-redshift galaxy overdensity, denoted by $\delta_{{\rm g}}$,
as being biased by the square of the relative velocity normalized
by its mean value, following Yoo et al. (2011). We then use perturbation
theory to compute the correlation function and 3PCF (\S4 \& \S6 respectively).

Writing the relative velocity as $\vec{v}_{{\rm bc}}=\vec{v}_{{\rm b}}-\vec{v}_{{\rm c}}$
(baryon velocity minus dark matter velocity) and $\sigma_{{\rm bc}}=\left<|\vec{v}_{{\rm bc}}|^{2}\right>^{1/2}$ (the root mean square value),
we define the dimensionless $v_{{\rm s}}^{2}=|\vec{v}_{{\rm bc}}|^{2}/\sigma_{{\rm bc}}^{2}$
and expand the galaxy overdensity in $v_{{\rm s}}^{2}-1$ to ensure
$\left<\delta_{{\rm g}}\right>=0$. Mathematically, this is
\begin{equation}
\delta_{{\rm g}}\left(\vec{r}\right)=\delta_{{\rm g},b_{v}=0}\left(\vec{r}\right)+b_{v}\left[v_{{\rm s}}^{2}\left(\vec{r}\right)-1\right],
\end{equation}
where 
\begin{equation}
\delta_{{\rm g},b_{v}=0}\left(\vec{r}\right)=b_{1}\delta_{{\rm m}}\left(\vec{r}\right)+b_{2}\left[\delta_{{\rm m}}^{2}\left(\vec{r}\right)-\left<\delta_{{\rm m}}^{2}\right>\right]
\end{equation}
captures the standard perturbation theory linear and non-linear bias
in the galaxy overdensity.\footnote{Often the non-linear bias coefficient is written as $b_2/2$ (e.g. Yoo et al. 2011; Yoo \& Seljak 2013), so care must be exercised in comparing values across different works.} 
$b_{v}$ in equation (1) is an unknown bias coefficient that, as discussed
in $\S1$, encodes how strongly the relative velocity affects galaxy
formation. $\delta_{{\rm m}}$ is the matter overdensity. 

We next outline how we compute the spatial structure of the relative
velocity, a required input in our bias model (1). Since the true
primordial density field at a given location is not known $a$ $priori$, we need to be able to compute the $v_s^2$ generated by an arbitrary density field.\footnote{We do require linear perturbation theory to be valid, so the field cannot be completely arbitrary.}  We therefore find the relative velocity due to a point perturbation
and then integrate it against the true density field. Though this latter is not known $a$ $priori$, its statistical properties are.  As we will only be considering expectation values over $v_s^2$, and thus over the density field, this is sufficient. 

Since the response
to an impulse is called the Green's function, we denote the relative
velocity due to a point perturbation by $\vec{v}_{{\rm G}}=v_{{\rm G}}\hat{r}$.
By symmetry, it must always point radially outward from the density
point sourcing it. It points outward because DM infalls under gravity
and baryons are static or pushed outwards by radiation pressure.

We now define the Green's function implicitly: 
\begin{equation}
\vec{v}_{{\rm bc}}\left(\vec{r},z\right)=\int v_{{\rm G}}\left(r_{1},z\right)\delta_{{\rm pri}}\left(\vec{r}+\vec{r}_{1}\right)\hat{r}_{1}d^{3}\vec{r}_{1}.
\end{equation}
To linear order, the relative velocity at redshift $z$ due to a primordial
density field $\delta_{{\rm pri}}\left(\vec{r}\right)$ is found by
integrating $\delta_{{\rm pri}}$ against the Green's function. Isotropy
demands that $\left<\vec{v}_{{\rm bc}}\right>=0$. Notice that $\vec{v}_{{\rm bc}}$
represents the dipole moment of the density field weighted by $v_{{\rm G}}$,
suggesting that multipole expansions will be natural moving forwards.
Figure 1 portrays schematically the use of the Green's function to
compute the relative velocity (and its square) due to an arbitary
density field.

The Green's function formalism above makes it evident that in our bias model, the square of the relative velocity contributes to the correlation function only beginning at
fourth order in the perturbed quantities. One overdensity is required
to source a relative velocity field, as shown in the lefthand panel of Figure 1, and to produce $v_{{\rm bc}}^{2}$
(equivalently, $v_{{\rm s}}^{2}),$ two overdensities are needed, as shown in the righthand panel of Figure 1.
For a Gaussian random field, all odd moments vanish, meaning the velocity
contributes to the correlation function beginning only at fourth order.
To obtain all of the fourth order contributions to the correlation function, we must expand $\delta_{{\rm m}}$ to second order in equation (2):
\begin{equation}
\delta_{{\rm m}}\left(\vec{r}\right)=\delta\left(\vec{r}\right)+\delta^{\left(2\right)}\left(\vec{r}\right).
\end{equation}
Here and throughout, $\delta$ is the linear density field while $\delta^{\left(2\right)}$
is the second-order density field, which is $\mathcal{O}\left(\delta^{2}\right)$.

\begin{figure}
\includegraphics[scale=0.4]{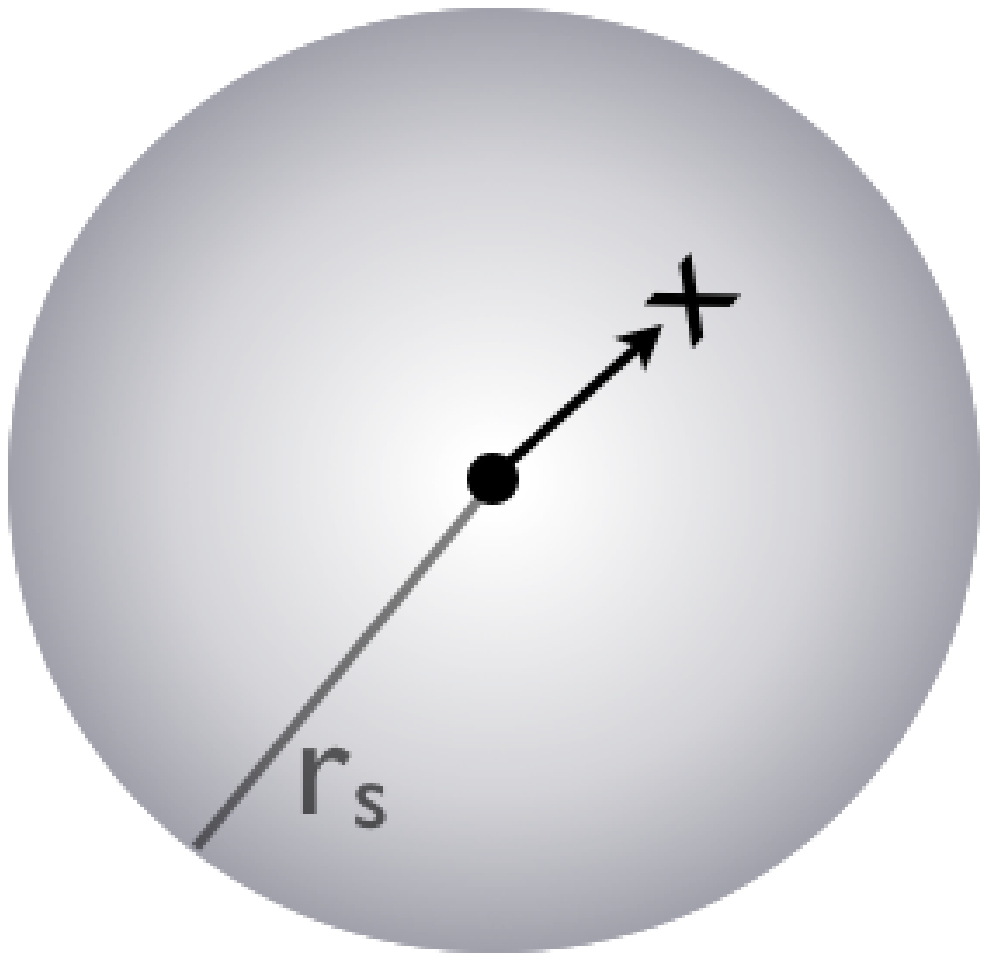}\includegraphics[scale=0.4]{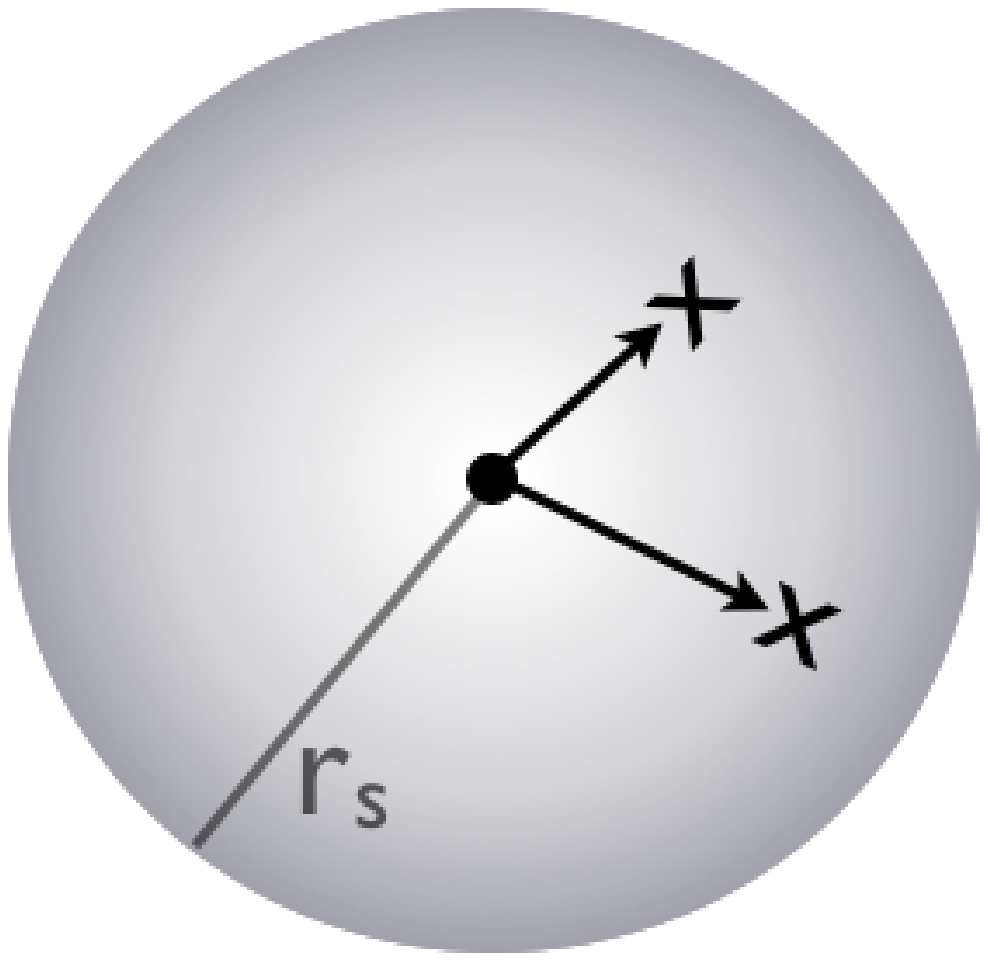}\caption{
Illustrations of the computation of the relative velocity field
and its square norm.  The \textbf{X} is a ``dummy'' density point to be integrated over. The left panel shows how $\vec{v}_{{\rm bc}}$ is evaluated at the dot by integrating the density field over all space weighted by the Green's function $v_{{\rm G}}.$
The darker shading indicates that $r^{2}v_{{\rm G}}$ peaks at $r_{{\rm s}}$,
a result discussed further in Figure 2. $\vec{v}_{{\rm bc}}^{2}$
(right panel) is evaluated analogously but using two copies of the
density field each weighted by the Green's function. In practice, because of the Green's function's structure (see Figure 2), only density points within $\sim r_{\rm s}$ of the dot significantly contribute to the relative velocity there.}
\end{figure}

Finally, we close this section with four further points about our perturbation theory framework. First, we note a subtlety of our bias model. $\left<v_{s}^{2}\right>=1$
and so if we carried through the expansion of the galaxy overdensity
to higher orders in $v_{s}^{2},$ we would expect these terms to contribute
to the correlation function at order unity times some combinatoric
factor. For instance, in the limit that $\xi\to\delta_{{\rm D}}^{\left[3\right]}$ (a 3-D Dirac delta function)
one can compute explicitly that a term $\left<v_{s}^{4}\left(0\right)\delta^{2}\left(\vec{r}\right)\right>\approx4$
appears in $\xi_{{\rm gg}}\left(\vec{r}\right)$ if equation (1) is
taken out to $v_{s}^{4}$. Since there are potentially an arbitrary
number of these terms, one might ask if our expansion converges.

However, physically, it is likely that the dimensionless parameter
of importance for galaxy formation is $|\vec{v}_{{\rm bc}}|^{2}/\sigma_{{\rm g}}^{2},$
where $\sigma_{{\rm g}}$ is some unknown, redshift-dependent circular
velocity or velocity dispersion for a typical galaxy. We expect that
$\left<|\vec{v}_{{\rm bc}}|^{2}/\sigma_{{\rm g}}^{2}\right>\ll1,$
so that an expansion in powers of this quantity would converge. Our
expansion, now with the coefficient of $v_s^{2n}$ labeled by $b_{vn}$, may be rewritten
in terms of $|\vec{v}_{{\rm bc}}|^{2}/\sigma_{{\rm g}}^{2}$ by taking 
\begin{equation}
b_{vn}v_s^{2n}= b_{vn\sigma_{g}}\left(\frac{v_{\rm bc}}{\sigma_{\rm g}}\right)^{2n}.
\end{equation}
The $b_{vn\sigma_{\rm g}}$ are coefficients of an expansion
in terms of powers of $|\vec{v}_{{\rm bc}}|^{2}/\sigma_{{\rm g}}^{2}$
and are assumed to be all intrinsically on the same order of magnitude.
Solving for $b_{vn}$ shows that it must fall rapidly with $n$ and our expansion
converges.

Second, we justify the use of linear perturbation theory. Though perturbation theory does not provide highly accurate
fits to simulation results on small scales ($\lesssim20\;{\rm Mpc}$),
the large scales ($\sim150\;{\rm Mpc}$) relevant for the BAO have
remained roughly linear down to the present day (for discussion of
non-linear effects, see Smith et al. 2003; Seo et al. 2008; Sherwin \&
Zaldarriaga 2012, though see also Roukema et al. 2014). The primary effect of what non-linear evolution
has occurred is to broaden the BAO peak in the galaxy-galaxy correlation
function, not to shift its center. As Eisenstein et al. (2007a)
show, the peak position is robust in configuration space. Further,
modern BAO surveys (e.g. Anderson et al. 2014) use reconstruction
to compute the peculiar velocity field implied by a given density
field and reverse it, thus allowing analysis to be performed on a
density field that is linear to even better approximation (Eisenstein
et al. 2007b; Seo et al. 2008). These considerations justify
our use of linear perturbation theory to compute how the relative velocity effect shifts the
BAO peak. It is unambiguous to calculate the lowest order change
in the correlation function and 3PCF the velocity produces. As we discuss above, higher order corrections
should quickly become negligible. One can debate the precise details
of the no-velocity correlation function, but the addition from the
velocity to any model chosen can be accurately computed in perturbation
theory. 

Third, when computing the expansion of the density field to second
order, we only consider effects generated by gravity.  For example, we
neglect effects due to couplings of radiation and matter. Naoz \& Barkana (2005) point out that on small scales, the sound speed
varies spatially after recombination due to density-dependent Compton
heating (see also Naoz et al. (2011)). This will not affect our conclusions because the BAO scale is
dominated by the relativistic sound speed prior to decoupling.  Another effect generated by coupling of radiation and matter is the impact of inhomogeneities in the intergalactic medium on the Lyman-$\alpha$ emission observed from galaxies (Wyithe \& Djikstra 2011).  This can create an additional clustering signal on large scales that would need to be properly accounted for if one wished to use the techniques in this work on a Lyman-$\alpha$ selected galaxy sample such as might be found using the Hobby-Eberly Telescope Dark Energy Experiment (HETDEX).

Finally, we consider the effects of redshift-space distortions (see
Hamilton 1998 for a review).  Peculiar velocities systematically
alter galaxy clustering even on large scales (Kaiser 1987; Bernardeau et al. 2002), introducing a strong directional dependence.  These distortions
do not substantially alter the Green's function picture, because
galaxy positions in redshift space are shifted by much less than
the acoustic scale.  However, the distortions can alter the resulting
correlations because the true large-scale correlations are also
small.  These effects can be accurately treated in cosmological
simulations, and we expect that studies of observational data would
want to compare to full simulations.  However, we note that
our analysis will average over triangles irrespective of their
orientation to the line of sight. While not optimal as regards
information content, such averages do tend to reduce the effects of
redshift distortions on large scales.  For example, the redshift-space
spherically averaged two-point correlation function on large scales
is primarily a rescaling of the real-space result, with a mild extra
broadening of the acoustic peak.  We similarly expect that our
orientation-averaged 3PCF results will be only mildly changed by
redshift-space distortions. Furthermore, the reconstruction of the linear density field discussed above can also correct redshift-space distortions on the scales relevant for this work by introducing a factor $1+f$, where $f=d(\ln D)/d(\ln a)$, $D$ is the linear growth function, and $a$ is the scale factor.  This factor represents the additional squashing along the line-of-sight (Eisenstein et al. 2007b).  This technique should allow removal of the redshift-space distortions on the large scales most significant for the signature presented here.

\section{Deriving the Green's function}

We now seek to obtain an explicit expression for the Green's function. We begin with
the linear theory continuity equation in configuration space. $a$
is the scale factor and we use comoving positions and velocities.
An overdot denotes a derivative with respect to time. We have
\begin{equation}
a^{-1}\nabla\cdot\vec{v}\left(\vec{r}\right)=-\dot{\delta}\left(\vec{r}\right),
\end{equation}
which Fourier transforms to
\begin{equation}
-ia^{-1}\vec{k}\cdot\tilde{\vec{v}}(\vec{k})=-\dot{\tilde{\delta}}(\vec{k}),
\end{equation}
where a tilde denotes a 3-D Fourier transform given by
\begin{equation}
\tilde{f}(\vec{k})=\int d^{3}\vec{r}f\left(\vec{r}\right)e^{i\vec{k}\cdot\vec{r}}
\end{equation}
with inverse transform
\begin{equation}
f\left(\vec{r}\right)=\int\frac{d^{3}\vec{k}}{\left(2\pi\right)^{3}}\tilde{f}(\vec{k})e^{-i\vec{k}\cdot\vec{r}}.
\end{equation}
For growing modes, $\tilde{\vec{v}}$ is parallel to $\vec{k}$ so
\begin{equation}
\tilde{\vec{v}}(\vec{k})=-\frac{i\vec{k}}{k^{2}}H\left(z\right)\frac{\partial\tilde{\delta}(\vec{k})}{\partial z}.
\end{equation}
This means
\begin{equation}
\tilde{\vec{v}}_{{\rm bc}}(\vec{k},z)=-\frac{i\vec{k}}{k^{2}}H\left(z\right)\frac{\partial(\tilde{\delta}_{{\rm b}}-\tilde{\delta}_{{\rm c}})}{\partial z}=-iT_{{\rm vbc}}\left(k,z\right)\tilde{\delta}_{{\rm pri}}(\vec{k})\hat{k},
\end{equation}
where subscript c is for CDM, b for baryons, vbc for relative velocity,
and we define the relative velocity transfer function
\begin{equation}
T_{{\rm vbc}}\left(k,z\right)=\frac{H\left(z\right)}{k}\frac{\partial}{\partial z}\left[T_{{\rm b}}\left(k,z\right)-T_{{\rm c}}\left(k,z\right)\right].
\end{equation}
$T_{{\rm b}}$ and $T_{{\rm c}}$ are the baryon and CDM transfer
functions, which give the evolution of each mode with redshift via
\begin{equation}
\tilde{\delta}_{{\rm b}}(\vec{k},z)=T_{{\rm b}}\left(k,z\right)\tilde{\delta}_{{\rm pri}}(\vec{k}),\;\tilde{\delta}_{{\rm c}}(\vec{k},z)=T_{{\rm c}}\left(k,z\right)\tilde{\delta}_{{\rm pri}}(\vec{k}).
\end{equation}
$\tilde{\delta}_{{\rm pri}}$ is the primordial density perturbation
related to the primordial power spectrum $P_{\rm pri}$ by $\left<\tilde{\delta}_{{\rm pri}}\left(\vec{k}\right)\tilde{\delta}_{{\rm pri}}^{*}\left(\vec{k}'\right)\right>=\left(2\pi\right)^{3}\delta_{{\rm D}}^{\left[3\right]}\left(\vec{k}-\vec{k'}\right)P_{{\rm pri}}\left(k\right)$, with $P_{{\rm pri}}=Ak^{n_{s}}$.\footnote{$A$ is fixed by the value of $\sigma_8$ today.} We emphasize
that the relative velocity transfer function maps the primordial density
field to a velocity field at some redshift $z,$ so $T_{{\rm vbc}}$
always acts on $\tilde{\delta}_{{\rm pri}}$. 

We now obtain the configuration space Green's function, defined implicitly
by equation (3). Using the Fourier representation of $\vec{v}_{{\rm bc}}$
(11) we have
\begin{align}
\vec{v}_{{\rm bc}}\left(\vec{r},z\right)&=\int\frac{d^{3}\vec{k}}{\left(2\pi\right)^{3}}e^{-i\vec{k}\cdot\vec{r}}\left[-iT_{{\rm vbc}}\left(k,z\right)\tilde{\delta}_{{\rm pri}}\left(\vec{k}\right)\hat{k}\right]\\
&=\int d^{3}\vec{r}_{1}\hat{r}_{1}v_{{\rm G}}\left(r_{1},z\right)\delta_{{\rm pri}}\left(\vec{r}+\vec{r}_{1}\right).\nonumber
\end{align}

We then rewrite $\tilde{\delta}_{{\rm pri}}\left(\vec{k}\right)=\int d^{3}\vec qe^{i\vec{k}\cdot\vec q}\delta_{{\rm pri}}\left(\vec{q}\right)$
to find
\begin{align}
&\int d^{3}\vec q\delta_{{\rm pri}}\left(\vec q\right)\int\frac{d^{3}\vec{k}}{\left(2\pi\right)^{3}}e^{i\vec{k}\cdot\vec q-i\vec{k}\cdot\vec{r}}\left[-iT_{{\rm vbc}}\left(k,z\right)\hat{k}\right]\\
&=\int d^{3}\vec{r}_{1}\hat{r}_{1}v_{{\rm G}}\left(r_{1},z\right)\delta_{{\rm pri}}\left(\vec{r}+\vec{r}_{1}\right).\nonumber
\end{align}
Changing variables on the left-hand side via $\vec q=\vec{r}-\vec{r}_{1}$
and then equating the resulting integrands over $d^{3}\vec{r}_{1}$, we have 
\begin{equation}
v_{{\rm G}}\left(r_{1},z\right)\hat{r}_{1}=\int\frac{d^{3}\vec{k}}{\left(2\pi\right)^{3}}e^{-i\vec{k}\cdot\vec{r}_{1}}\left[iT_{{\rm vbc}}\left(k,z\right)\hat{k}\right],
\end{equation}
which, projecting onto $\hat{r}_{1}$, results in 
\begin{align}
v_{{\rm G}}\left(r_{1},z\right)&=\int\frac{k^{2}dk}{2\pi^{2}}\int_{-1}^{1}\frac{d\mu}{2}\mu e^{-ikr_{1}\mu}\left[iT_{{\rm vbc}}\left(k,z\right)\right]\\
&=\int\frac{k^{2}dk}{2\pi^{2}}j_{1}\left(kr_{1}\right)T_{{\rm vbc}}\left(k,z\right).\nonumber
\end{align}
 Above, $\mu=\hat{k}\cdot\hat{r}_1$ and $j_1$ is the spherical Bessel function of order one.  We now have the desired velocity Green's function.  Noting that its Fourier transform is closely related to the velocity transfer function, for notational consistency we define
\begin{equation}
\tilde{v}_{{\rm G}}\left(k,z\right)=\int4\pi r^{2}drj_{1}\left(kr\right)v_{{\rm G}}\left(r,z\right)=T_{{\rm vbc}}\left(k,z\right)
\end{equation}
and use $\tilde{v}_{\rm G}$ going forward.

In practice, we compute $v_{\rm G}$ by transforming $\tilde{v}_{\rm G}$ using equation (17). Thus we must first compute $\tilde{v}_{{\rm G}}$. Using a flat $\Lambda CDM$
cosmology with $\Omega_{{\rm b}}h^{2}=0.0226$, $\Omega_{{\rm c}}h^{2}=0.112$, $n_s=0.96$,
and $H_{0}=70\;{\rm km/s/Mpc}$, we output transfer functions $T_{{\rm b}}$
and $T_{{\rm c}}$ from CAMB (Lewis 2000) on a grid equally spaced
in $\log k$ with 5,000 divisions per decade from $k=6.95\times 10^{-5}$ to $10.50$. To approximate $\partial T_{{\rm b}}/\partial z$ and $\partial T_{{\rm c}}/\partial z$ (cf. equation
(12)), we discretize the derivative at each redshift $z$ with $\Delta z=0.10z$. To avoid ringing due to the finiteness of our grid
in Fourier space, we use a smoothing $\exp\left[-k^{2}\right]$ to
evaluate the integral (17) (and all analogous integrals over $dk$
in what follows).

Figure 2 shows the Green's functions for $v_{{\rm b}},\; v_{{\rm c}},$
and $v_{{\rm bc}}$ at $z\sim1020$. We have multiplied each by $r^{2}$ for two reasons.
First, for a random distribution of densities, a spherical shell of
radius $r$ will contribute as $r^{2}$ when integrated over volume.
Second, this weighting renders the fine structure more apparent.
The most striking feature of Figure 2 is the compact support of the $v_{{\rm bc}}$ Green's
function. This occurs because for $r>r_{{\rm s}}$,
$v_{{\rm bc}}\to0$, as radiation pressure cannot support the baryons
against gravitational infall. Also salient is that $v_{{\rm bc}}\approx v_{{\rm c}}$
for $r\lesssim0.9r_{{\rm s}}$: baryons are in hydrostatic equilibrium
with $v_{{\rm b}}\approx0$. There is a bump in the baryon velocity at $r_{s}$ due to
the outgoing baryon-photon overdensity from the BAO.  Meanwhile, the DM
infalls. 

Inside $r_{{\rm s}},$ the DM infalls as roughly $\sim1/r$ rather
than $1/r^{2}$ because the baryon-photon fluid's contribution to
the potential, which, during radiation domination, overshadows that of the DM overdensity at the
origin, is diluted. For a test particle at $r<r_{\rm s}$,  some of the baryon-photon
overdensity is outside a Gaussian sphere of radius $r$, and hence does not contribute to the gravitational force felt by the particle. This is
equivalent to the fact that modes inside the horizon grow less quickly
than those outside the horizon during radiation domination, and is
why the DM transfer function turns over for $k\sim k_{{\rm eq}},$
$k_{{\rm eq}}$ the wavenumber entering the horizon at matter-radiation
equality. Using the continuity equation in Fourier space (equation
(7)), slower growth of a given mode implies a lower velocity field
for that mode, so modes inside the horizon indeed have a lower velocity field
than those outside the horizon.

\begin{figure}
\centering\includegraphics[width=0.6\textwidth]{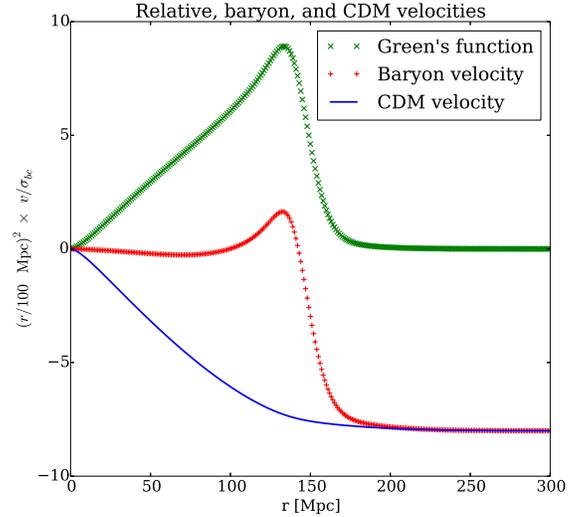}\caption{Relative velocity Green's function $v_{{\rm G}}$, the relative velocity $v_{{\rm bc}}=v_{{\rm b}}-v_{{\rm c}}$
due to a $\delta_{{\rm D}}^{\left(3\right)}$ density perturbation
at the origin. We show the Green's function at $z\sim1020,$ as at
this epoch the relative velocity freezes in (cf. \S1). Here and throughout the paper,
green will denote the Green's function, plot symbol X will pertain
to functions involving the relative velocity, red will be baryons,
and blue CDM. Note that for $r>r_{s}\simeq150\;{\rm Mpc},$ baryons and
DM infall at the same speed. In contrast, inside $r_{s}$ the baryons
are mostly locked in hydrostatic equilibrium while the DM infalls
roughly as $1/r$ due to the dilution of the radiation density perturbation
as this latter expands with time (see \S3 for further discussion). The slight bump in the baryon velocity represents baryons in motion as the fluid compresses and expands at the sound horizon. The neutrinos, which travel at roughly $c$ and so are outside $r_{\rm s}$, smooth the transition to zero $v_{\rm bc}$ outside the sound horizon, as does Silk damping (Silk 1968).}
\end{figure}

\section{Analysis of the two-point correlation function}

\subsection{The shift in $\xi_{\rm gg}$}

We now wish to compute $\xi_{v}$, the relative velocity contribution
to the correlation function. We define $\xi_{{\rm gg}}$ as the full
galaxy-galaxy correlation function including $\xi_{v}$ and denote
the late-time linear matter correlation function $\xi$. To compute
$\xi_{v}$, we use the galaxy overdensity bias model of $\S2$. Numerical
subscripts denote spatial positions: $\delta_{1}\equiv\delta(\vec{r}_{1})$.
For more compact notation, we also define $\delta_{v}=v_{{\rm s}}^{2}-1$;
note this is second-order in $\delta$. We have the velocity contributions
to the product $\delta_{{\rm g}}\left(\vec{r}_{1}\right)\delta_{{\rm g}}\left(\vec{r}_{2}\right)$:
\begin{align}
&\left[\delta_{{\rm g}}\left(\vec{r}_{1}\right)\delta_{{\rm g}}\left(\vec{r}_{2}\right)\right]_{v}=b_{1}b_{v}\left(\delta_{{\rm m}1}\delta_{v2}+\delta_{{\rm m}2}\delta_{v1}\right)+b_{2}b_{v}\\
&\times\left(\delta_{{\rm m}1}^{2}\delta_{v2}+\delta_{{\rm m}2}^{2}\delta_{v1}-\left<\delta_{{\rm m}}^{2}\right>\left[\delta_{v1}+\delta_{v2}\right]\right)+b_{v}^{2}\delta_{v1}\delta_{v2}.\nonumber
\end{align}
Defining $r=\big|\vec{r}_{2}-\vec{r}_{1}\big|$ and noting that terms in $\delta_{1}\delta_{v2}$ vanish because we assume
a Gaussian random field, we have
\begin{align}
&\xi_{v}\left(r\right) \equiv\left<\left[\delta_{{\rm g}}\left(\vec{r}_{1}\right)\delta_{{\rm g}}\left(\vec{r}_{2}\right)\right]_{v}\right>= \xi_{v1}+\xi_{v2}+\xi_{vv}\\
&\equiv 2b_{1}b_{v}\left<\delta_{1}^{(2)}\delta_{v2}\right>
+2b_{2}b_{v}\left[\left<\delta_{1}^{2}\delta_{v2}\right>-\left<\delta_{v}\right>\left<\delta^{2}\right>\right]\nonumber
+b_{v}^{2}\left<\delta_{v1}\delta_{v2}\right>.\nonumber
\end{align}
 The factors of 2 come from invoking homogeneity so that $\left<\delta_{1}^{(2)}\delta_{v2}\right>=\left<\delta_{2}^{(2)}\delta_{v1}\right>$
and $\left<\delta_{1}^{2}\delta_{v2}\right>=\left<\delta_{2}^{2}\delta_{v1}\right>.$
Moving forward we will often denote the term proportional to $b_{1}$ by $\xi_{v1}$
and analogously for the terms in $b_{2}$ and $b_{v}$, as indicated above. We have also dropped terms above fourth order.

Using the definition of $\delta_{v}$, we can simplify the three terms in $\xi_{v}$ to:
\begin{equation}
\xi_{v1}\left(r\right)\equiv2b_{1}b_{v}\left[\left<\delta_{1}^{(2)}v_{s2}^{2}\right>-\left<\delta^{(2)}\right>\right],
\end{equation}
\begin{equation}
\xi_{v2}\left(r\right)\equiv2b_{2}b_{v}\left[\left<\delta_{1}^{2}v_{s2}^{2}\right>-\xi\left(0\right)\right],
\end{equation}
\begin{equation}
\xi_{vv}\left(r\right)\equiv b_{v}^{2}\left[\left<v_{s1}^{2}v_{s2}^{2}\right>-1\right].
\end{equation}
For equation (22), we have replaced $\left<\delta^{2}\right>$ by $\xi\left(0\right)$. These three terms (ignoring the constant offsets) are shown schematically in Figure 3.   Spheres indicate a velocity squared, while the solid square shows the density field squared and the solid triangle represents the second-order density field $\delta ^{(2)}$. The dotted lines show the correlations that remain after the constants above are subtracted off.

Note that these expectation values involve products of four values
of the linear density field at different locations.  Since the linear density field is a Gaussian Random Field, we can use
Wick's theorem to simplify.  The constants subtracted above are just generated
by the presence of $\left<\delta_{{\rm m}}^{2}\right>$ in our model
for $\delta_{{\rm g}}$, and ultimately cancel when Wick's theorem
is applied.

Finally, we need to compute $v_{s}^{2}$ since it enters equations (21)-(23).  We have
\begin{align}
v_{s}^{2}\left(\vec{r}\right)&=\frac{1}{\sigma_{{\rm bc}}^{2}\left(z\right)}\int d^{3}\vec{r}_{1}d^{3}\vec{r}_{2}v_{{\rm G}}\left(r_{1},z\right)v_{{\rm G}}\left(r_{2},z\right)\\
&\times\delta_{{\rm pri}}\left(\vec{r}_{1}+\vec{r}\right)\delta_{{\rm pri}}\left(\vec{r}_{2}+\vec{r}\right)\hat{r}_{1}\cdot\hat{r}_{2},\nonumber
\end{align}
with $v_{{\rm G}}$ given by equation (17). See the right panel of Figure 1 for a visual reminder of how $v_s^2$ is calculated from the Green's function.  For $v_s^2$, we in turn require $\sigma_{{\rm bc}}^{2}\equiv\left<\vec{v}_{{\rm bc}}\cdot\vec{v}_{{\rm bc}}\right>$, which can be found by writing $\vec{v}_{{\rm bc}}\left(\vec{r}\right)$
in the Fourier basis using equations (14) and (15), squaring, and
taking expectation values. One then uses the relation between $\tilde{\delta}_{\rm pri}$ and $P_{\rm pri}$ given in \S3 to simplify the integrals and finds
\begin{equation}
\sigma_{{\rm bc}}^{2}\left(z\right)=\int\frac{k^{2}dk}{2\pi^{2}}\tilde{v}_{{\rm G}}^{2}\left(k,z\right)P_{{\rm pri}}\left(k\right),
\end{equation}
in agreement with Dalal et al. (2010).

Note that after decoupling, the relative velocity is no longer
sourced and will therefore decay as $a^{-1}$, as will ${\rm \sigma_{{\rm bc}}}.$
Hence after decoupling $v_{s}^{2}\left(\vec{r}\right)$ is redshift-independent.

\begin{figure}
\includegraphics[scale=0.35]{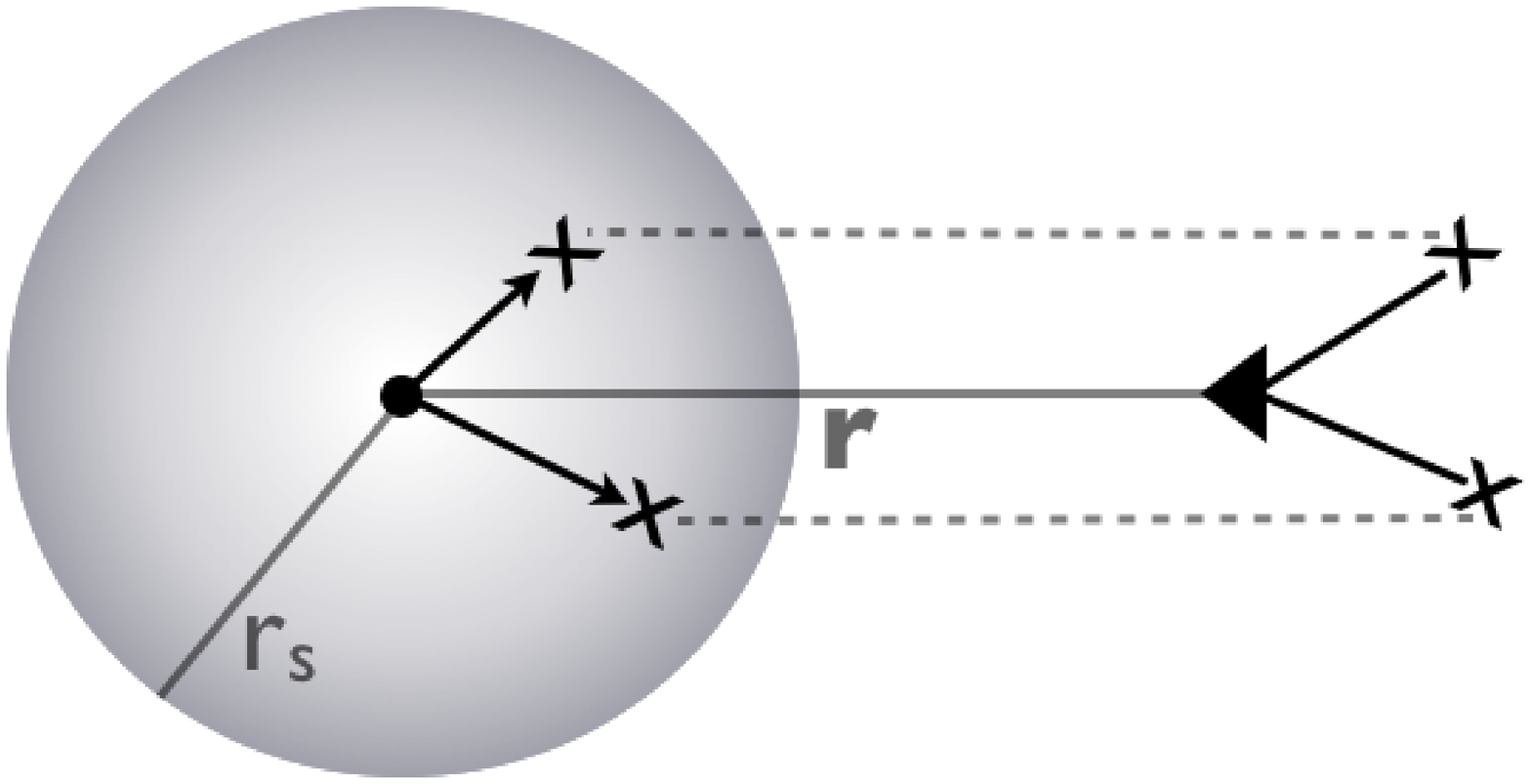}

\includegraphics[scale=0.4]{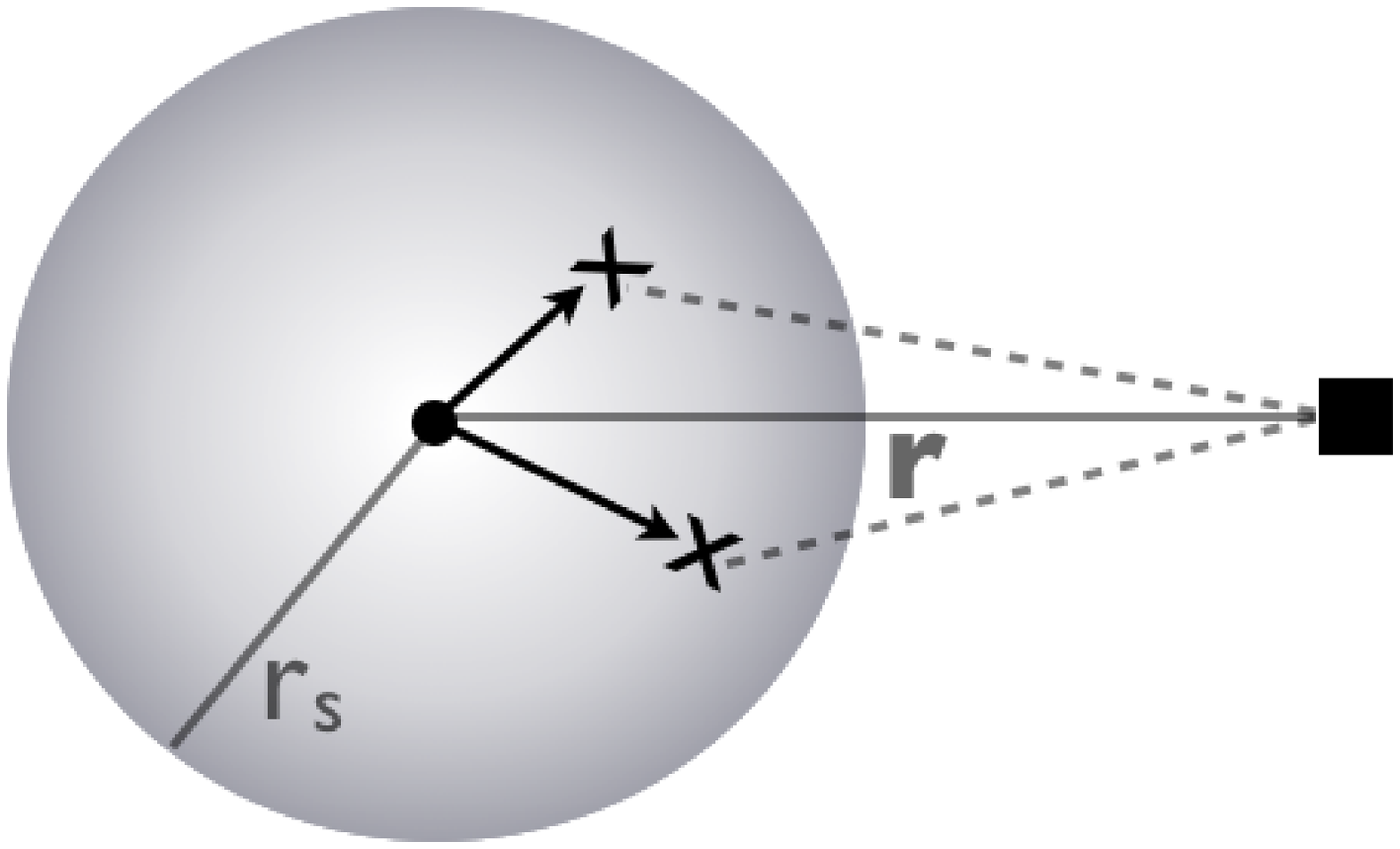}

\includegraphics[scale=0.35]{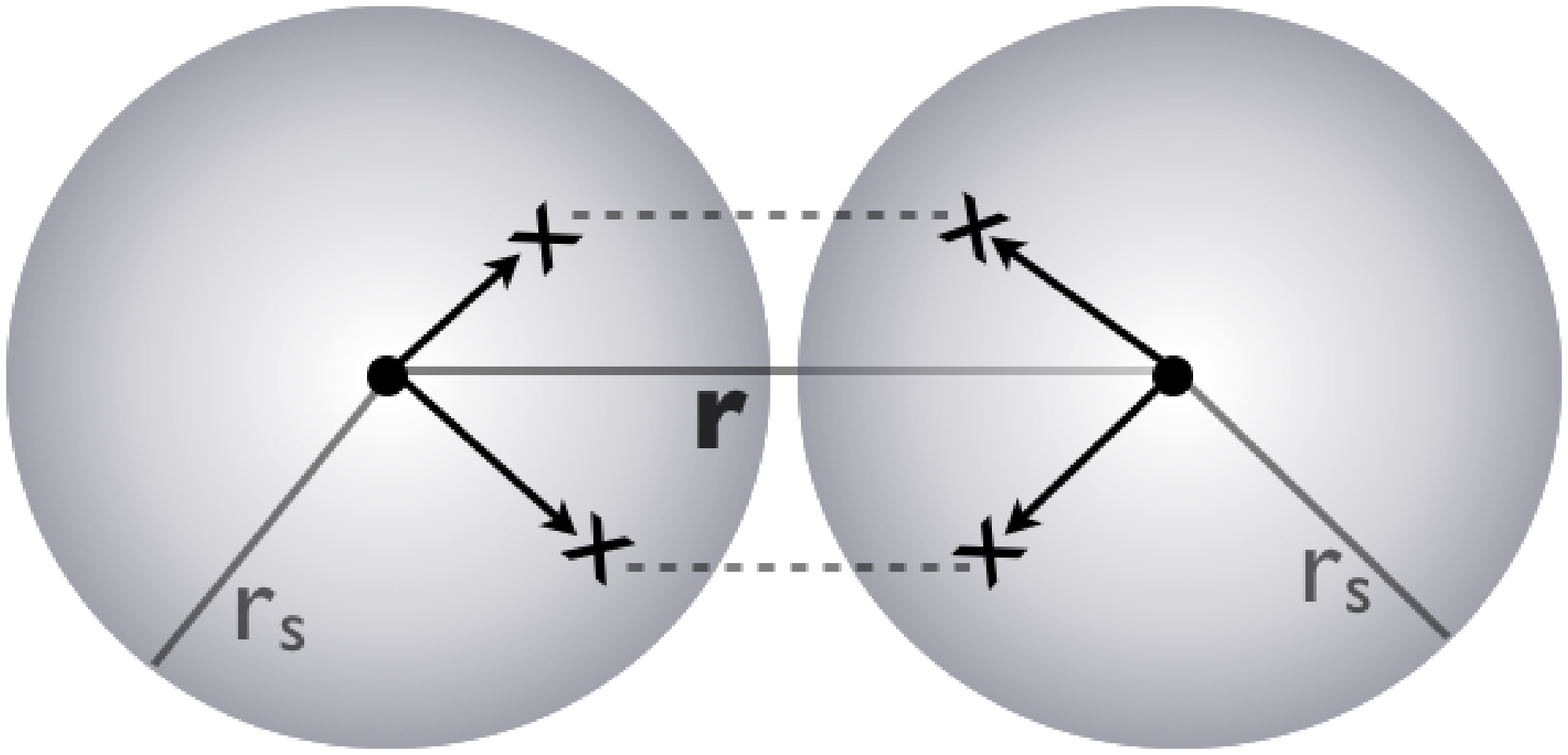}\caption{Illustrations of the three contributions to $\xi_v$.  The top panel is $\xi_{v1}\left(r\right)$ (equation (21)), the middle panel $\xi_{v2}\left(r\right)$ (equation (22)), and the bottom panel $\xi_{vv}\left(r\right)$
(equation (23)). The symbols are as explained in Figure 1; the dotted
lines show the correlations that remain after simplifying using Wick's
theorem. The triangle (top panel) is where $\delta^{\left(2\right)}\left(\vec{r}\right)$
is evaluated, by integrating over the entire surrounding density field, while the square (middle panel) is where $\delta^2\left(\vec{r}\right)$ is evaluated. These diagrams make the convolutional
structure of this calculation evident; it is to be further discussed in \S4.2.}
\end{figure}

\subsection{$\xi_v$ as convolutions}

We now show that $\xi_{v}$ can be expressed as convolutions of various
functions of the linear density field with a 6-dimensional velocity
kernel we define below. This latter preserves the radial structure
of the Green's function, and this insight will support our analysis
in \S5 of why the BAO peak shifts.

We begin with $\xi_{v2}$ (equation (22)) because it is the simplest. The disconnected
part of $\left<\delta_{1}^{2}v_{s2}^{2}\right>$ is $\left<\delta_{1}^{2}\right>\left<v_{s2}^{2}\right>=\xi\left(0\right)$
and so cancels off. Focusing on the remaining terms from Wick's theorem,
represented by the dotted lines in Figure 3, we find
\begin{equation}
\xi_{v2}\left(r\right)=4b_{2}b_{v}\int_{34}\xi_{\times}\left(\big|\vec{r}-\vec{r}_{3}\big|\right)\xi_{\times}\left(\big|\vec{r}-\vec{r}_{4}\big|\right)V_{3,4},
\end{equation}
where the factor of $2$ relative to equation (22) comes from Wick's theorem. To connect with Figure 3, note that we have put the origin at the black dot (i.e. used homogeneity to set $\vec{r}_2=0$ in equation (22)).
We have defined the 6-dimensional velocity kernel
\begin{equation}
V_{a,b}\equiv v_{G}\left(r_{a}\right)v_{G}\left(r_{b}\right)\left(\hat{r}_{a}\cdot\hat{r}_{b}\right)/\sigma_{{\rm bc}}^{2}.
\end{equation}
This kernel generates $v_{s}^{2}\left(0\right)$ from integrals over
$\delta_{{\rm pri,}a}$ and $\delta_{{\rm pri,}b}$ for all $\vec{r}_{a}$
and $\vec{r}_{b}$ within roughly $r_{{\rm s}}$ of the origin. Note that
it is redshift-independent. We have also defined $\xi_{\times}$: a heterosynchronous
correlation of the low-redshift linear theory matter density field
with the primordial density field:
\begin{equation}
\xi_{\times}\left(\vec{r}\right)=\int\frac{d^{3}\vec{k}}{\left(2\pi\right)^{3}}e^{-i\vec{k}\cdot\vec{r}}P_{\times}
\end{equation}
where
\begin{equation}
P_{\times}\equiv Ak^{n_{s}}T_{{\rm m}}\left(k,z=0\right).
\end{equation}
The matter transfer function is $T_{{\rm m}}=(\Omega_{{\rm c}}T_{{\rm c}}+\Omega_{{\rm b}}T_{{\rm b}})/(\Omega_{{\rm c}}+\Omega_{{\rm b}})$.
Note that $P_{\times}$ is not the standard power spectrum, which would use the square of the transfer function. Rather, is is a cross power spectrum between the low-redshift linear theory matter
field (we use $z=0$) and the primordial density field.

We define the 6-D convolution as
\begin{equation}
\left[f_{1,2}\star g_{1,2}\right]\left(\vec{R}\right)=\int d^{3}\vec{r}_{1}d^{3}\vec{r}_{2}f\left(\vec{r}_{1},\vec{r_{2}}\right)g\left(\vec{r}_{a}-\vec{r}_{1},\vec{r}_{b}-\vec{r}_{2}\right),
\end{equation}
where $\vec{R}=\left(\vec{r}_{a},\vec{r}_{b}\right)$ is a 6-D vector
and $\vec{r}_{a}$ and $\vec{r}_{b}$ are 3-D vectors. With this definition,
it is immediate to write
\begin{equation}
\xi_{v2}\left(r\right)=4b_{2}b_{v}\left[V_{3,4}\star\xi_{\times3}\xi_{\times4}\right]\left(\vec{r},\vec{r}\right).
\end{equation}
We now show that $\xi_{vv}$ (equation (23)) can also be expressed as a convolution.
Canceling the disconnected part and then evaluating $\xi_{vv}$ (see
Figure 3), we find
\begin{align}
\xi_{vv}\left(r\right)&=2b_{v}^{2}\int_{3456}\xi_{{\rm pri}}\left(\big|\vec{r}+\vec{r}_{3}-\vec{r}_{5}\big|\right)\xi_{{\rm pri}}\left(\big|\vec{r}+\vec{r}_{4}-\vec{r}_{6}\big|\right)\nonumber\\
&\times V_{3,4}V_{5,6},
\end{align}
where again the factor of $2$ relative to equation (23) is from Wick's theorem.
Note that for $\xi_{vv}=b_{v}^{2}\left<\delta_{v1}\delta_{v2}\right>$
we are correlating a velocity field with a velocity field, and so
we must use only $\delta_{{\rm pri}}$, leading to $\xi_{{\rm pri}}=\left(2\pi\right)^{-3}\int d^{3}\vec{k}e^{-i\vec{k}\cdot\vec{r}}P_{{\rm pri}}$
in the expression above. Let us first consider the integrals over
$d^{3}\vec{r}_{3}d^{3}\vec{r}_{4}$ above. Flipping the signs of $\vec{r}_{3}$
and $\vec{r}_{4,}$ which leaves the Jacobian unchanged, and applying
the definition (30), we obtain
\begin{equation}
\int_{34}\xi_{{\rm pri}}\left(\big|\vec{r}+\vec{r}_{3}\big|\right)\xi_{{\rm pri}}\left(\big|\vec{r}+\vec{r}_{4}\big|\right)V_{3,4}=\left[V_{34}\star\xi_{{\rm pri}3}\xi_{{\rm pri}4}\right](\vec{r},\vec{r}).
\end{equation}
Inserting this in equation (32), we have
\begin{equation}
\xi_{vv}\left(r\right)=2b_{v}^{2}\int_{56}V_{5,6}\left[V_{34}\star\xi_{{\rm pri}3}\xi_{{\rm pri}4}\right](\vec{r}-\vec{r}_{5},\vec{r}-\vec{r}_{6}).
\end{equation}
Applying again the definition (30), we find
\begin{equation}
\xi_{vv}\left(r\right)=2b_{v}^{2}\left[V_{5,6}\star\left[V_{3,4}\star\xi_{{\rm pri}3}\xi_{{\rm pri}4}\right]\left(\vec{r}_{5},\vec{r}_{6}\right)\right]\left(\vec{r},\vec{r}\right).
\end{equation}

Finally, we turn to $\xi_{v1}$ (equation (21)), the most difficult term to evaluate
because it involves the second-order density field $\delta^{\left(2\right)}$.
Using the same procedure as for $\xi_{v2}$ and $\xi_{vv}$, we have 
\begin{align}
\xi_{v1}\left(r\right)&=4b_{1}b_{v}\int_{3456}\xi_{\times}\left(\big|\vec{r}+\vec{r}_{5}-\vec{r}_{3}\big|\right)\xi_{\times}\left(\big|\vec{r}+\vec{r}_{6}-\vec{r}_{4}\big|\right)\nonumber\\
&\times F_{3,4}^{\left(2\right)}V_{5,6},
\end{align}
where the factor of $2$ relative to equation (21) is again from Wick's theorem. Again, Figure 3 illustrates the relevant correlations.
$F_{3,4}^{(2)}$ is the Fourier transform of the second-order kernel
\begin{align}
&\tilde{F}^{\left(2\right)}\left(\vec{k}_{1},\vec{k}_{2}\right)=\frac{5}{7}+\frac{1}{2}\hat{k}_{1}\cdot\hat{k}_{2}\left(\frac{k_{1}}{k_{2}}+\frac{k_{2}}{k_{1}}\right)+\frac{2}{7}\left(\hat{k}_{1}\cdot\hat{k}_{2}\right)^{2}\nonumber\\
&=\frac{17}{21}P_{0}\left(\hat{k}_{1}\cdot\hat{k}_{2}\right)+\frac{1}{2}\left(\frac{k_{1}}{k_{2}}+\frac{k_{2}}{k_{1}}\right)P_{1}\left(\hat{k}_{1}\cdot\hat{k}_{2}\right)\\
&+\frac{4}{21}P_{2}\left(\hat{k}_{1}\cdot\hat{k}_{2}\right)\nonumber
\end{align}
(Goroff et al. 1986; Jain \& Bertschinger 1994; Bernardeau et al. 2002 {[}equation (45){]}).
Analogously to $V_{3,4}$, an integral of two density fields $\delta_{3}$
and $\delta_{4}$ against $F_{3,4}^{(2)}$ generates a second-order
density field $\delta^{(2)}(0)$. The $P_i$s are Legendre polynomials.

Working now in Fourier space, we see that equation (36) becomes
\begin{align}
\xi_{v1}\left(r\right) &=4b_{1}b_{v}\bigg\{ \bigg[\int\frac{d^{3}\vec{k}_{1}d^{3}\vec{k}_{2}}{\left(2\pi\right)^{6}}\tilde{F}^{(2)}\left(\vec{k}_{1},\vec{k}_{2}\right)P_{\times}\left(k_{1}\right)\\
&\times P_{\times}\left(k_{2}\right)e^{-i\vec{k_{1}}\cdot\vec{r}_{3}}e^{-i\vec{k}_{2}\cdot\vec{r}_{4}}\bigg]\left(\vec{r}_{3},\vec{r}_{4}\right)\star V_{5,6}\bigg\} \left(\vec{r},\vec{r}\right).\nonumber
\end{align}
Note that the integral over $d^{3}\vec{r}_{3}d^{3}\vec{r}_{4}$ in
equation (36) is the 6-D convolution $\xi_{\times}\left(\vec{r}\right)\xi_{\times}(\vec{r})\star F_{3,4}^{(2)}$,
which we have rewritten as a product in Fourier space using the Convolution
Theorem to obtain equation (38). In equation (38), the integral in
square brackets becomes a function of $ $$\vec{r}_{3}$ and $\vec{r}_{4}$
when evaluated; this in turn is convolved with $V_{5,6}$. For convenience,
we define the integral in square brackets as 
\begin{align}
f_{\times3,4}&=\bigg[\int\frac{d^{3}\vec{k}_{1}d^{3}\vec{k}_{2}}{\left(2\pi\right)^{6}}\tilde{F}^{(2)}\left(\vec{k}_{1},\vec{k}_{2}\right)P_{\times}\left(k_{1}\right)\\
& \times P_{\times}\left(k_{2}\right)e^{-i\vec{k_{1}}\cdot\vec{r}_{3}}e^{-i\vec{k}_{2}\cdot\vec{r}_{4}}\bigg]\left(\vec{r}_{3},\vec{r}_{4}\right)\nonumber.
\end{align}
It is more convenient to work with this representation than to directly
consider $\xi_{\times}\left(\vec{r}\right)\xi_{\times}(\vec{r})\star F_{3,4}^{(2)}$,
since $F_{3,4}^{(2)}$ (the Fourier transform of $\tilde{F}^{(2)}$)
is divergent without the regularization multiplication of $\tilde{F}^{(2)}$ by the smoothed cross
power spectrum provides. With this notation, we now have
\begin{equation}
\xi_{v1}(r)=4b_{1}b_{v}\left[f_{\times3,4}\star V_{3,4}\right]\left(\vec{r},\vec{r}\right).
\end{equation}
Thus, from equations (31), (35), and (40), we see that all three terms
in $\xi_{v}$ are just convolutions of various functions built from
the correlation function with the 6-D velocity kernel (27) we have defined.
This latter ultimately encodes the radial structure of the Green's
function, shown in Figure 2. 

We now pause to examine the limit that $\xi\to\delta_{{\rm D}}^{\left[3\right]}$,
as this will offer strong physical intuition for the behavior we should
see once we have numerically evaluated the equations above. In this
limit, $\xi_{v2}\left(r\right)=4b_{2}b_{v}v_{G}^{2}\left(r\right)$,
while $\xi_{vv}\left(r\right)=2b_{v}^{2}\left[V_{1,2}\star V_{1,2}\right]\left(\vec{r}\right)$
(unfortunately $\xi_{v1}$ is divergent in this limit).

The approximate form that $v_{{\rm G}}^{2}\propto1/r^{2}$ for $r\leq r_{{\rm s}}$
(see Figure 2) means that inside $r_{{\rm s}},$ each radial bin of
volume $dV=4\pi r^{2}dr$ will give an equal contribution to $\xi_{v2}$.
We thus expect that $r^{2}\xi_{v2}$ will be approximately a step
function, constant for $r\leq r_{{\rm s}}$ and zero otherwise. Meanwhile,
$V_{1,2}\star V_{1,2}$ is an autocorrelation, so $\xi_{vv}$ will
peak at $\vec{r}=0.$ A second, lesser peak in the autocorrelation
occurs at $|\vec{r}|=r_{{\rm s}}$, which becomes the most prominent
one when we consider the volume-weighted $4\pi r^{2}\xi_{vv}\left(r\right)$.
Therefore $r^{2}\xi_{vv}$ should have a well-defined peak that encodes
the acoustic scale, and have support out to $\sim2r_{{\rm s}}$. These
behaviors are displayed in the lower panel of Figure 5,
magenta dashed and orange X-ed curves.

\subsection{Evaluating the convolutions}
We now give details on how the convolutions of the previous
section can be quickly evaluated. Formally,
the convolutions are multidimensional integrals, and so could be computed directly
via Monte Carlo methods. We avoid this by showing that in principle all of the convolutions can be analytically reduced to
one-dimensional radial integrals, because the angular dependences
of all functions involved are known. This is highly desirable and
may be done using results we prove in the Appendix. 

First, we explicitly obtain $f_{\times3,4}$; since this is also useful
in computing the 3PCF, we write it down here. We have
\begin{align}
&f_{\times3,4} =\mathcal{H}_{0}^{-1}\left\{ \frac{17}{21}P_{\times}\left(k_{1}\right)P_{{\rm }\times}\left(k_{2}\right)\right\} (r_{3},r_{4})P_{0}\left(\hat{r}_{3}\cdot\hat{r}_{4}\right)\\
&-\mathcal{H}_{1}^{-1}\left\{ \frac{1}{2}\left(\frac{k_{1}}{k_{2}}+\frac{k_{2}}{k_{1}}\right)P_{\times}\left(k_{1}\right)P_{\times}\left(k_{2}\right)\right\} (r_{3},r_{4})P_{1}\left(\hat{r}_{3}\cdot\hat{r}_{4}\right)\nonumber\\
&+\mathcal{H}_{2}^{-1}\left\{ \frac{4}{21}P_{\times}\left(k_{1}\right)P_{\times}\left(k_{2}\right)\right\} (r_{3},r_{4})P_{2}\left(\hat{r}_{3}\cdot\hat{r}_{4}\right),\nonumber
\end{align}
where the $\mathcal{H}^{-1}s$, defined in the Appendix (equations
(61) and (62)), are 2-D transforms composed of 1-D integrals against
spherical Bessel functions (these latter integrals are closely related to Hankel
transforms). Simplifying (see Appendix for formulae used to do so), we have
\begin{align}
&f_{{\rm }\times3,4}=\frac{17}{21}P_{0}\left(\hat{r}_{3}\cdot\hat{r}_{4}\right)\xi_{\times3}\xi_{\times4}-\frac{1}{2}P_{1}\left(\hat{r}_{3}\cdot\hat{r}_{4}\right)\\
&\times \left[\xi_{{\rm }\times4}^{[1-]}\xi_{\times3}^{[1+]}+\xi_{\times4}^{[1+]}\xi_{{\rm }\times3}^{[1-]}\right]
+\frac{4}{21}P_{2}\left(\hat{r}_{3}\cdot\hat{r}_{4}\right)\xi_{{\rm }\times3}^{[2]}\xi_{{\rm }\times4}^{[2]},\nonumber
\end{align}
with
\begin{equation}
\xi_{{\rm }\times}^{[1\pm]}\left(r\right)\equiv\int_{0}^{\infty}\frac{k^{2}dk}{2\pi^{2}}j_{1}\left(kr\right)P_{\times}\left(k\right)k^{\pm1}
\end{equation}
and
\begin{equation}
\xi_{\times}^{[2]}\left(r\right)\equiv\int_{0}^{\infty}\frac{k^{2}dk}{2\pi^{2}}j_{2}\left(kr\right)P_{\times}\left(k\right).
\end{equation}
This reduces the terms in $\xi_{v}$ to
\begin{align}
&\xi_{v1}\left(r\right)=4b_{1}b_{v}\sigma_{{\rm bc}}^{-2}\bigg(\frac{31}{35}\left[h_{1}^{-1}\left\{ \tilde{v}_{{\rm G}}P_{\times}\right\} (r)\right]^{2}\\
&-\frac{1}{3}\bigg[h_{0}^{-1}\left\{ \tilde{v}_{{\rm G}}P_{{\rm }\times}k\right\} (r)h_{0}^{-1}\left\{ \tilde{v}_{{\rm G}}P_{{\rm }\times}k^{-1}\right\} (r)\nonumber\\
&+2h_{2}^{-1}\left\{ \tilde{v}_{{\rm G}}P_{\times}k\right\} (r)h_{2}^{-1}\left\{ \tilde{v}_{{\rm G}}P_{{\rm }\times}k^{-1}\right\} (r)\bigg]\nonumber\\
&+\frac{12}{105}\left[h_{3}^{-1}\left\{ \tilde{v}_{{\rm G}}P_{{\rm }\times}\right\} (r)\right]^{2}\bigg),\nonumber
\end{align}
\begin{equation}
\xi_{v2}\left(r\right)=4b_{2}b_{v}\sigma_{{\rm bc}}^{-2}\left[h_{1}^{-1}\left\{ \tilde{v}_{{\rm G}}P_{{\rm }\times}\right\} (r)\right]^{2},
\end{equation}
and
\begin{align}
\xi_{vv}\left(r\right)&=\frac{2}{3}b_{v}^{2}\sigma_{{\rm bc}}^{-4}\bigg[\left[h_{0}^{-1}\left\{ \tilde{v}_{{\rm G}}^{2}P_{{\rm pri}}\right\} (r)\right]^{2}\\
&+2\left[h_{2}^{-1}\left\{ \tilde{v}_{{\rm G}}^{2}P_{{\rm {\rm pri}}}\right\} (r)\right]^{2}\bigg].\nonumber
\end{align}
Note that the right-hand side of each equation is a function of $r$,
as $\tilde{v}_{{\rm G}}$ and $P_{\times}$ (equations (18) and (29)) are functions of $k$,
and the $h_{l}^{-1}$s bring them to functions of $r$ (cf. equation
(62)). The equations above allow efficient numerical evaluation of
$\xi_{v}$. Finally, it should be noted that Wick's theorem yields
immediately that $\xi_{vv}\left(0\right)$ $=\left(2/3\right)b_{v}^{2}$ (cf. equation (23)).
Taking this limit of equation (47) explicitly cross-checks our use
of the convolutional formalism:
\begin{equation}
\xi_{vv}\left(0\right)=\frac{2}{3}b_{v}^{2}\sigma_{{\rm bc}}^{-4}\left[\int\frac{k^{2}dk}{2\pi^{2}}\tilde{v}_{{\rm G}}^{2}P_{{\rm pri}}\right]^{2}=\frac{2}{3}b_{v}^{2},
\end{equation}
with the last equality using equation (25).

\section{Explaining the BAO peak shift}

We begin with a descriptive sketch of why the BAO peak
shifts and then move to more detailed discussion of our numerical
results. First: the key aspect of $\xi_{v}$ that allows it to shift
the BAO peak in the correlation function is the sharp drop to zero
for $r$ greater than roughly $r_{s}$. This can be traced back to
the same feature in the relative velocity Green's function (see Figure
2). Thus $\xi_{v}$ adds $\left(b_{v}>0\right)$ or subtracts $\left(b_{v}<0\right)$
probability density from $\xi_{{\rm gg}}$ asymmetrically, doing so
primarily {\it inwards} of $r_{{\rm s}}$. It is this asymmetric alteration
that pulls the BAO peak in or out in scale. Fundamentally, then, the
shift is due to the presence of an acoustic horizon: the relative
velocity is at root a pressure effect, and as such can alter correlations
roughly only within the sound horizon. 

In greater mathematical detail, the 6-D velocity kernel $V_{1,2}$ has support only for $r\lesssim r_{s}$.
Convolving $\xi_{{\rm \times}}$ and $f_{\times}$ with it act
as smoothing operations that simply broaden its peak. Since the smoothing functions are so narrow, the convolutions are only non-zero essentially where the velocity kernel is non-zero. For $b_{v}>0,$
this adds probability density inwards of $r_{{\rm s}}$, while for $b_{v}<0$, this subtracts probability
density inwards of $r_{{\rm s}}$. Figure 4 shows the two smoothing functions, $\xi_{{\rm \times}}$
and $f_{\times}$ (built on transforms of $\xi_{{\rm }}$), and the
$v_{{\rm bc}}$ Green's function they smooth. The behavior of the contribution from $\xi_{vv}$ is complex, as it is a double convolution, but as we learned from the $\xi_{\times}\to\delta_{\rm D}^{[3]}$ limit, it has support out to $\sim 2r_{\rm s}$.  However, around the BAO peak location $\xi_{vv}$ is roughly symmetric, so it does not contribute much to shifting the peak (Figure 5).

We now move to discuss each term in $\xi_v$ more extensively; each term is shown in Figure 5.  Two of the terms are fairly simple in structure, and can be understood by recalling the limit $\xi\to\delta_{\rm D}^{[3]}$ of \S4.2.  $\xi_{v2}$ is roughly constant for $r<r_{\rm s}$, and drops quickly to zero outside the sound horizon.  It extends slightly beyond the sound horizon; this may be traced back to the smooth drop of the velocity Green's function (Figure 2) produced by the neutrinos and Silk damping.  The small bump near the sound horizon is due to the baryons' velocity there; this has its origin in the baryon velocity's contribution to the Green's function (see Figure 2).  This is confirmed by the bottom panel of Figure 5, plotting the $\xi\to\delta_{\rm D}^{[3]}$ limit.  Here, a bump is still present near the acoustic scale, meaning that the bump is due to structure in the velocity Green's function and not due to any structure in $\xi_{\times}$. Since $\xi_{v2}$ is non-zero essentially solely inside $r_{\rm s}$, it adds asymmetrically to the correlation function and hence can pull the BAO bump inwards ($b_v>0$) or outwards ($b_v<0$).

$\xi_{vv}$ is also fairly simple and can also be understood by recalling the limit $\xi\to\delta_{\rm D}^{[3]}$ of \S4.2.  From equation (35), we see that it is a double convolution, and that the first convolution produces a  function identical, up to amplitude, to $\xi_{v2}$.  However, this is then convolved with an additional velocity kernel, which is, unlike $\xi_{\times}$ and $f_{\times}$, rather broad.  Thus there is overlap between the velocity kernel and the first convolution until nearly twice the sound horizon.  However, $\xi_{vv}$ is roughly symmetric around the acoustic scale out to $\sim 20\;\rm{Mpc}$ on either side of it; this means that in the region of the BAO peak, $\xi_{vv}$ is adding symmetrically to $\xi$ and hence will contribute minimally to shifting the peak.

We now discuss the last remaining term in $\xi_v$, $\xi_{v1}$.  This term has the most complicated structure, a consequence of its generation by the second-order density field.  This latter is non-local: computing $\delta^{(2)}$ requires integrating over all space (see Figure 3, top panel, and equation (36)).  Hence this term requires correlating the entire linear density field with the velocity field.  Nonetheless, its behavior can be understood qualitatively.  The second-order density field (or the $\tilde{F}_2$ kernel) represents non-linear gravitational evolution, which, in our Green's function approach, pulls mass towards the origin.  Thus it is not surprising that $f_{\times}$ is rather narrow. Since $f_{\times}$ is so narrow, it will only have non-zero overlap with $V_{3,4}$ where the latter is non-zero.  This explains the compact support of $\xi_{v1}$.  $\xi_{v1}$ is able to become negative because $f_{\times}$ does.  Note that (Figure 4) $f_{\times}$ is positive very close to the origin; this is what permits $\xi_{v1}$ to become positive at intermediate scales, $r\sim 50-150\;\rm{Mpc}$.  Overall, for $b_v>0$, $\xi_{v1}$ mostly adds to $\xi_v$ inwards of the BAO bump and subtracts from it directly outside the sound horizon; this asymmetric effect can shift the BAO peak.  

In summary, then, we have seen that of the three terms entering $\xi_v$, only two make significant contributions to a peak shift.  In both cases, the shift is due to the compact support of these terms, which in turn results from the narrowness of the smoothing functions combined with the compact support of the velocity Green's function.  We now consider the relative importance of these two terms when they are added together with plausible values of the linear and non-linear bias.  Note that both of these terms are proportional to $b_v$, so different values of $b_v$ will not alter their relative weights.  $\xi_{v1}$ is a factor of $\sim5-10$ larger than $\xi_{v2}$, and so it dominates the total velocity contribution to $\xi_{\rm gg}$ for $b_1=1,\;b_2=0.1$.  The total velocity contribution to $\xi_{\rm gg}$ in Figure 6 thus looks very similar to $\xi_{v1}$.  Note however that the effects of $\xi_{vv}$ are perceptible.  In particular, inside $r_{\rm s}$, the addition for $b_v>0$ is greater in magnitude than that for $b_v<0$. This is because for $b_v>0$ $\xi_{v1}$ and $\xi_{vv}$ are both mostly positive within $r_{\rm s}$, while for $b_v<0$, $\xi_{v1}$ flips sign but $\xi_{vv}$ does not (compare $b_v=-2\%$ and $b_v=2\%$ curves in Figure 6).  Outside $r_{\rm s}$, the case is reversed.  For $b_v<0$, $\xi_{v1}>0$ as is $\xi_{vv}$, while for $b_v>0$, these two terms have opposite signs and thus interfere destructively.  This point explains why the curve for $b_v=-2\%$ is larger in magnitude than that for $b_v=2\%$ outside $r_{\rm s}$. Finally, the convergence of the $b_{v}<0$ curves to those with $b_{v}>0$ at large $r$ is because the $b_{v}^{2}$ term is the only contribution for $r\gg r_{{\rm s}}$ and it is symmetric in $b_{v}$. Note also that the curves
are nearly anti-symmetric under sign flip in $b_{v}$, not surprising
since two of the terms manifestly have this symmetry. They are not
perfectly anti-symmetric under this transformation due to the $b_{v}^{2}$
term. 

It is significant that the $\xi_{v1}$ term dominates the velocity addition to $\xi_{\rm gg}$ because this means one need not measure $b_2$ as well as one measures $b_1$ to obtain a good correction to the correlation function.  Fortunately observationally $b_1$ is indeed measured better than $b_2$.

\begin{figure}
  \raggedleft
  \begin{minipage}{8cm}
  \includegraphics[width=11cm]{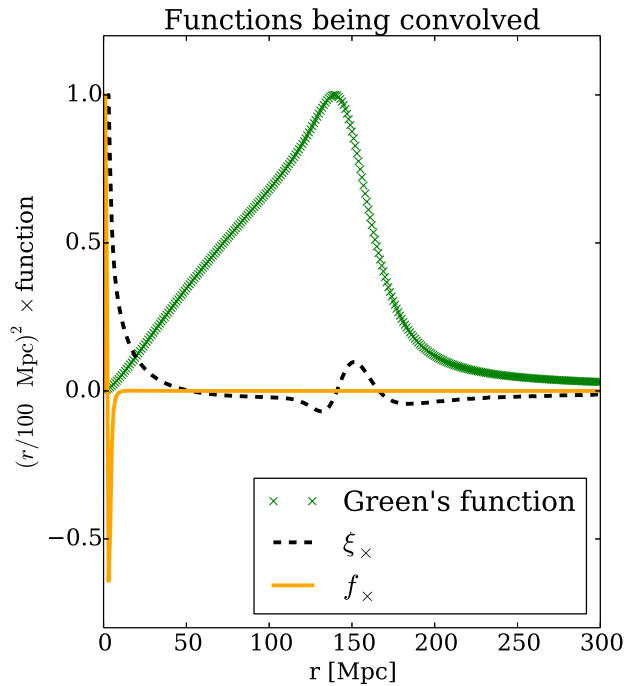}
\caption{Equations (31), (35), and (40) display the correction $\xi_{v}$ to
$\xi$ as the 6-D convolutions of 2 different 6-D kernels, $\xi_{\times3}\xi_{\times4}$
and $f_{\times3,4}$, with $V_{1,2}$. These kernels have angular
dependence, so the convolutions genuinely are fully 6-D. However,
we can gain heuristic intuition for their behavior by setting all
angle-dependent terms to unity and considering the radial behavior
of each function. Each function then becomes a product of two copies
of the same radial function at $r_{3}$ and $r_{4}$. We plot one
copy of each here, against $v_{{\rm G}}$, which appears in $V_{1,2}=v_{{\rm G1}}v_{{\rm G2}}\left(\hat{r}_{1}\cdot\hat{r}_{2}\right)$.
The two functions convolved with $v_{G}$ are both sharply peaked
at $r=0,$ so the convolution does not greatly alter $v_{G}$; therefore
we expect all results involving it to have distinctive structure at
the acoustic scale. As a reminder, $\xi_{\times}$ generates $\xi_{v2}$ and $f_{\times}$ generates $\xi_{v1}.\;\;\xi_{vv}$ is generated by convolving $V_{1,2}$ with itself; intuition may be gained for this case by imagining convolving two copies of the Green's function above.}
\end{minipage}
\end{figure}

\begin{figure}
\centering\includegraphics[width=0.47\textwidth]{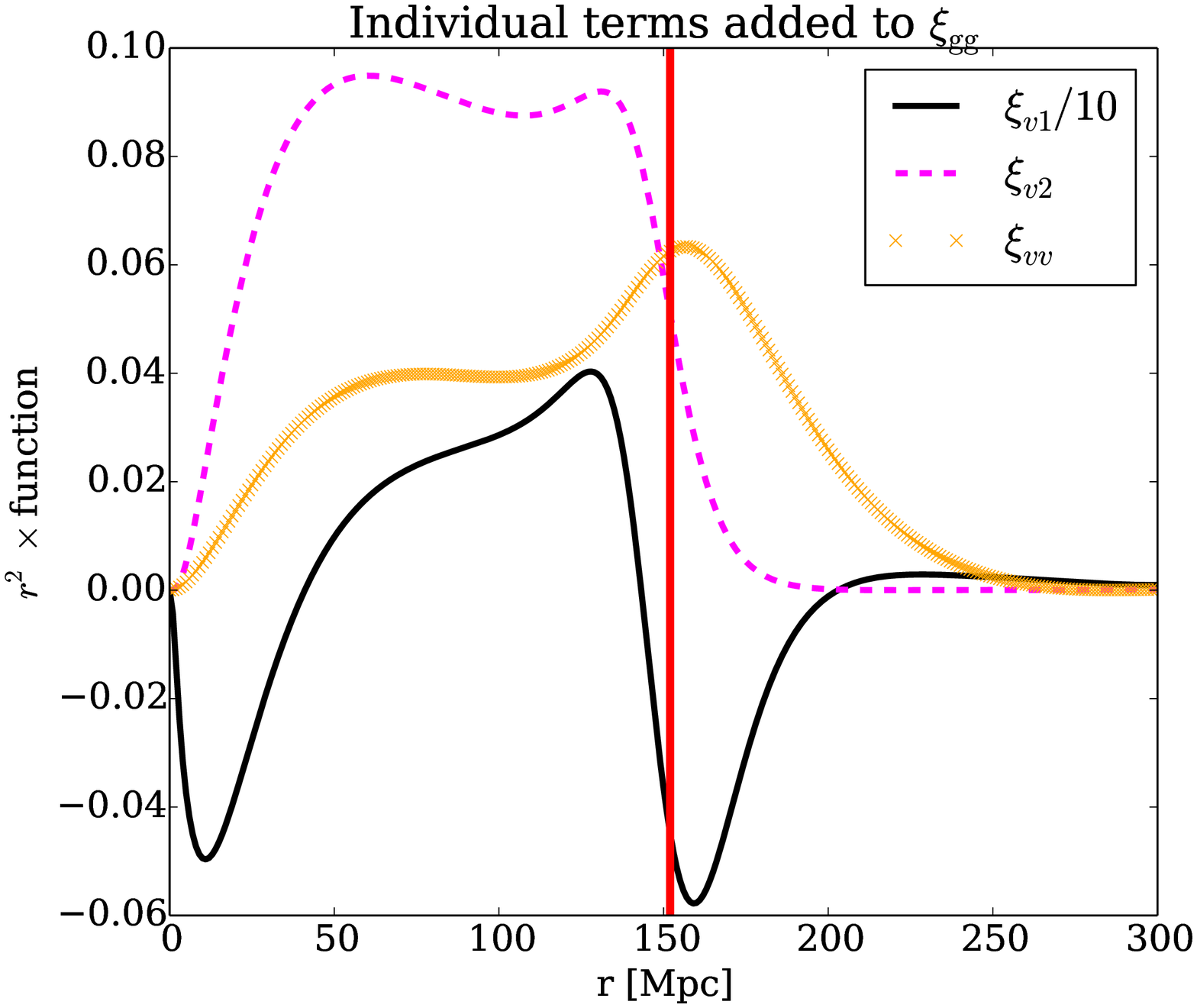}

\centering\includegraphics[width=0.47\textwidth]{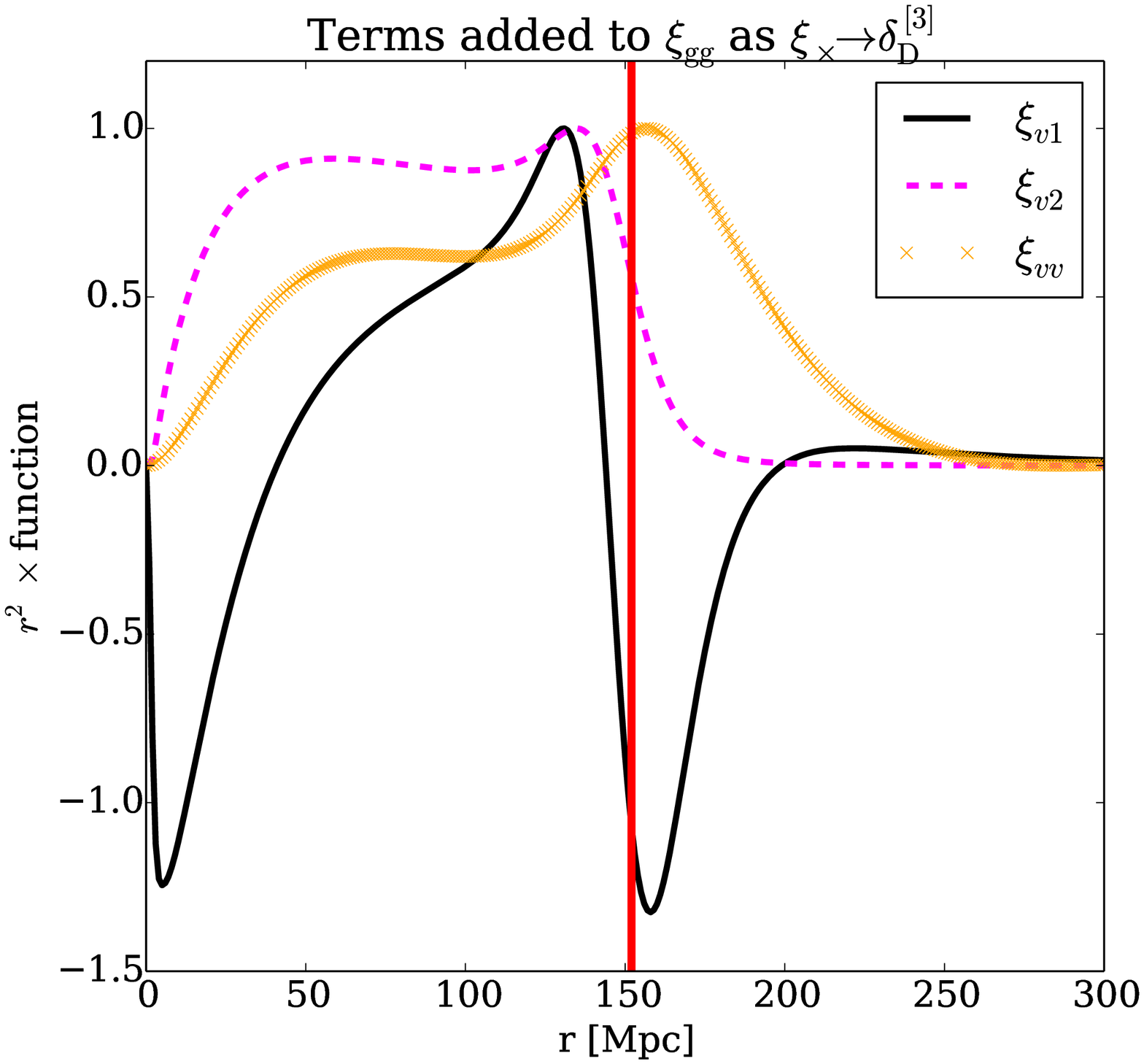}\caption{The top panel shows the three terms in $\xi_{v}$ (equations (21)-(23)) computed using
equations (45)-(47). We have used $b_{1}=1,b_{2}=0.1,b_{v}=0.01$. We have added a vertical red line at the BAO scale in the late-time linear correlation function, $150\;\rm{Mpc}$.
The bottom panel shows the three terms in $\xi_{v}$ in the limit $\xi_{\times}\to\delta_{{\rm D}}^{\left(3\right)}.$
Here we have scaled each term to have maximum $=1$, as the limit
that $\xi_{\times}\to\delta_{{\rm D}}^{\left[3\right]}$ requires
some normalization. Formally, the limit that $\xi_{\times}\to\delta_{D}^{[3]}$
means $P_{\times}\to1,$ but we use a smoothing such that $P\to\exp\left[-k^{2}\right]$
to avoid introducing ringing due to our finite grid in Fourier space.
This smoothing effectively regularizes the divergence of $f_{\times 3,4}$
discussed in \S4.2. The three curves are discussed in detail in \S5; note that (top panel) $\xi_{v1}$ dominates and has support only roughly within $r_{\rm s}$.  This is also true for $\xi_{v2}$.  While $\xi_{vv}$ has support out to $\sim2r_{\rm s}$, it adds to $\xi_{\rm gg}$ roughly symmetrically about the BAO peak location and so will not shift the peak much. Note the similarity of the two panels; this is because $\xi_{\times}$ is already very narrow compared to $v_{\rm G}$.  The $\xi_{\times}\to\delta_{\rm{D}}^{[3]}$ limit tells us that there is some intrinsic width to the peaks in $\xi_{v1},\;\xi_{v2},$ and $\xi_{vv}$.  For instance the trough in $\xi_{v1}$ around $r_{\rm s}$  is $\sim 15\;\rm{Mpc}$ wide in the bottom panel.  It widens to $\sim 40\;\rm{Mpc}$ in the top panel, consistent with the fact that $\xi_{\times}$ has about this width and so will smooth structure on this scale (see Figure 4).}
\end{figure}

\begin{figure}
\centering\includegraphics[width=0.47\textwidth]{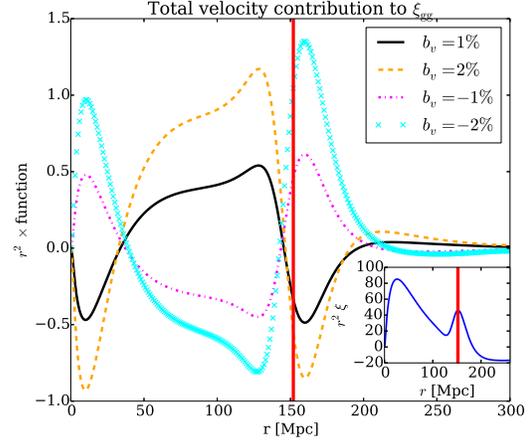}

\caption{$\xi_{v}$ (equation (20)) with different values of $b_{v}$. We have used $b_{1}=1,b_{2}=0.1.$  The case we trace throughout the paper is $b_v=1\%$, but even $b_v\sim 2\%$ is allowed by current constraints (Yoo \& Seljak 2013), hence its presentation here. We have marked the BAO scale in the late-time linear correlation function with a red line.  The inset is $\xi$, the linear-theory matter correlation function today, which we present to highlight the BAO peak that $\xi_v$ can shift. The key point of this plot is that $\xi_v$ alters the correlation function asymmetrically about $r_{\rm s}$, mostly altering it inwards of $r_{\rm s}$.  This creates a shift in the peak. A secondary point is the similarity with Figure 5, top panel, $\xi_{v1}$ (compare black solid curves between the two figures); see \S5 for further discussion. Finally note that the results are not exactly symmetric in the sign of $b_v$, especially visible at scales larger than $150\;\rm{Mpc}$ where the $b_v^2$ term dominates.}
\end{figure}

\section{Isolating $\lowercase{b}_{\lowercase{v}}$ from the 3PCF: pre-cyclic computation}

The three-point galaxy correlation function (3PCF) describes the excess probability over random of finding three galaxies with positions $\vec{r}_1$, $\vec{r}_2$, and $\vec{r}_3$; that is, in a given triangle configuration.  We compute it at fourth order using our bias model ((1), (2), and (4)) as
\begin{align}
&\zeta(\vec{r}_{1},\vec{r}_{2},\vec{r}_{3})\equiv<\delta_{{\rm g}_{1}}\delta_{{\rm g}_{2}}\delta_{{\rm g}_{3}}>=\zeta_{{\rm pc}}\left(\vec{r}_{3}-\vec{r}_{1},\vec{r}_{2}-\vec{r}_{1}\right)\\
&+\zeta_{{\rm pc}}\left(\vec{r}_{1}-\vec{r}_{2},\vec{r}_{3}-\vec{r}_{2}\right)+\zeta_{{\rm pc}}\left(\vec{r}_{2}-\vec{r}_{3},\vec{r}_{1}-\vec{r}_{3}\right)\nonumber.
\end{align}
We now rewrite the pre-cyclic piece (subscript ``pc'')  as a function of two side lengths and their enclosed angle (so that numerical subscripts denote sides of the triangle rather than spatial positions):
\begin{align}
&\zeta_{{\rm pc}}(r_{1},r_{2};\theta_{12})=b_{1}^{2}b_{v}\left[<\delta_{1}\delta_{2}\delta_{v_{3}}>-\xi_{12}\left<\delta_{v}\right>\right]+b_{1}^{2}b_{2}\\
&\times\left[<\delta_{1}\delta_{2}\delta_{3}^{2}>-\xi_{12}\left<\delta^{2}\right>\right]+b_{1}^{3}\left[<\delta_{1}\delta_{2}\delta_{3}^{(2)}>-\xi_{12}\left<\delta^{(2)}\right>\right].\nonumber
\end{align}
Here,
the phrase ``pre-cyclic'' means that we choose one vertex of the triangle
of galaxies at which to define $\theta_{12}$ and the two sides enclosing
it, $r_{1}$ and $r_{2}$. Note that in the product of three copies
of $\delta_{g}$ needed to form the 3PCF, each of the three galaxies
can contribute a $\delta_{v},$ a $\delta^{2}$ and a $\delta^{(2)}$.
The pre-cyclic term is written by choosing one galaxy to contribute
each of these (it may be the same one). We have chosen the third galaxy to
contribute all of these more complicated terms. Since we
can then take this galaxy to be at the origin, this approach simplifies
the calculation. However, to connect with observations, where there
is no ``preferred'' vertex (galaxy), we eventually must sum cyclically,
giving each galaxy in the survey the chance to contribute $\delta_{v},\;\delta^{2},$
and $\delta^{(2)}$. This procedure and its results are described
in \S7.

We now may calculate $\zeta_{{\rm pc}}$ explicitly from perturbation theory.\footnote{Notice that $\xi_{v1},\; \xi_{v2}$, and $\xi_{vv}$ (equations (40), (31), and (35)) are in fact just convolutions of $V_{3,4}$ with the three terms of $\zeta_{\rm pc}$. Unfortunately, the weights with which these convolutions enter  (respectively $4$, $4$, and $2$) differ from those with which the three terms of $\zeta_{\rm pc}$  enter $\zeta_{\rm pc}$ (equal weights), so $\xi_v$ is not quite the convolution of $V_{3,4}$ with $\zeta_{\rm pc}$.}
Motivated by the fact that $\vec{v}_{{\rm bc}}$ is a weighted dipole
moment of the density field, we expand the angular dependence of $\zeta_{{\rm pc}}$
using Legendre polynomials as a basis.\footnote{In the more general context of discussing a new set of estimators for 3-point statistics, many years before the discovery of the relative velocity effect, Szapudi (2004) suggested decomposing the bispectrum's angular dependence in Legendre polynomials. Our work here differs because we are interested in the configuration-space behavior as it is there that the acoustic structure of the relative velocity will be distinctive. Importantly, subsequent work of Pan and Szapudi (2005) exploits this basis in configuration space (a measurement of the monopole moment of the 3PCF in 2dFGRS), illustrating the utility of this decomposition in both Fourier and configuration space.}
We find
\begin{equation}
\zeta_{{\rm pc}}\left(r_1,r_2;\theta_{12}\right)=\sum_{l=0}^{2}\zeta_{{\rm pc}l}\left(r_{1},r_{2}\right)P_{l}\left(\cos\theta_{12}\right),
\end{equation}
with the coefficients as
\begin{equation}
\zeta_{{\rm pc}0}\left(r_{1},r_{2}\right)=\left[2b_{1}^{2}b_{2}+\frac{34}{21}b_{1}^{3}\right]\xi_{1}\xi_{2},
\end{equation}
\begin{align}
\zeta_{{\rm pc}1}\left(r_{1},r_{2}\right)&=-b_{1}^{3}\left[\xi_{1}^{\left[1-\right]}\xi_{2}^{\left[1+\right]}+\xi_{2}^{\left[1-\right]}\xi_{1}^{\left[1+\right]}\right]\\
&+2b_{1}^{2}b_{v}\left[V_{3,4}\star\xi_{\times3}\xi_{\times4}\right]\left(r_{1},r_{2}\right)\nonumber
\end{align}
and
\begin{equation}
\zeta_{{\rm pc}2}\left(r_{1},r_{2}\right)=\frac{8}{21}b_{1}^{3}\xi_{1}^{\left[2\right]}\xi_{2}^{\left[2\right]}.
\end{equation}
$\xi$ is the linear theory matter correlation function at $z=0$; $\xi^{\left[1\pm\right]}$ and 
$\xi^{\left[2\right]}$ are defined
analogously to the earlier $\xi_{\times}^{\left[1\pm\right]}$ and $\xi_{\times}^{\left[2\right]}$ (equations (43) and (44)) as:
\begin{equation}
\xi^{[1\pm]}\left(r\right)=\int_{0}^{\infty}\frac{dk}{2\pi^{2}}k^{2}j_{1}\left(kr\right)P\left(k\right)k^{\pm1}.
\end{equation}
and
\begin{equation}
\xi^{[2]}\left(r\right)=\int_{0}^{\infty}\frac{dk}{2\pi^{2}}k^{2}j_{2}\left(kr\right)P\left(k\right).
\end{equation}
Recall that numerical subscripts on $\xi$ and $V$ indicate spatial
position. Factors of 2 enter all terms above (e.g. $2b_{1}^{2}b_{2}$)
due to Wick's theorem, which reduces the four linear fields implict
in each expectation value of equation (50) to a sum of products of three expectation
values over two linear fields each, one of which cancels off due to the
subtractions in equation (50). The two that remain are the same by
homogeneity. The $b_1^3$ coefficients, $34/21$ in equation (52), $-1$ in equation (53),
and $8/21$ in equation (54) can be traced back to equation (37) for
the kernel generating the second-order density field, multiplied by
the $2$ from Wick's theorem. The negative sign in equation (53) relative to this kernel is because the $b_1^3$ term in equation (53) comes from a Legendre polynomial of odd order $(l=1)$ in equation (37) and so picks up a factor of $(-1)^l=-1$ when transformed according to equation (60).

The functions appearing in the pre-cyclic term, and the products of
these functions evaluated on isosceles triangles, are shown in Figure
7. In both plots the structure around $\sim r_{{\rm s}}$ should
particularly be noted, as it is ultimately why the 3PCF will have
signatures of both the standard BAO and the relative velocity. Finally, in the pre-cyclic 3PCF, $b_{v}$ enters only at $l=1$,
the dipole term. Given that it is this term that generates $\left<v_{{\rm bc}}^{2}\right>$
in the first place, this result is not surprising. 

\begin{figure}
\begin{center}
\includegraphics[width=0.57\textwidth]{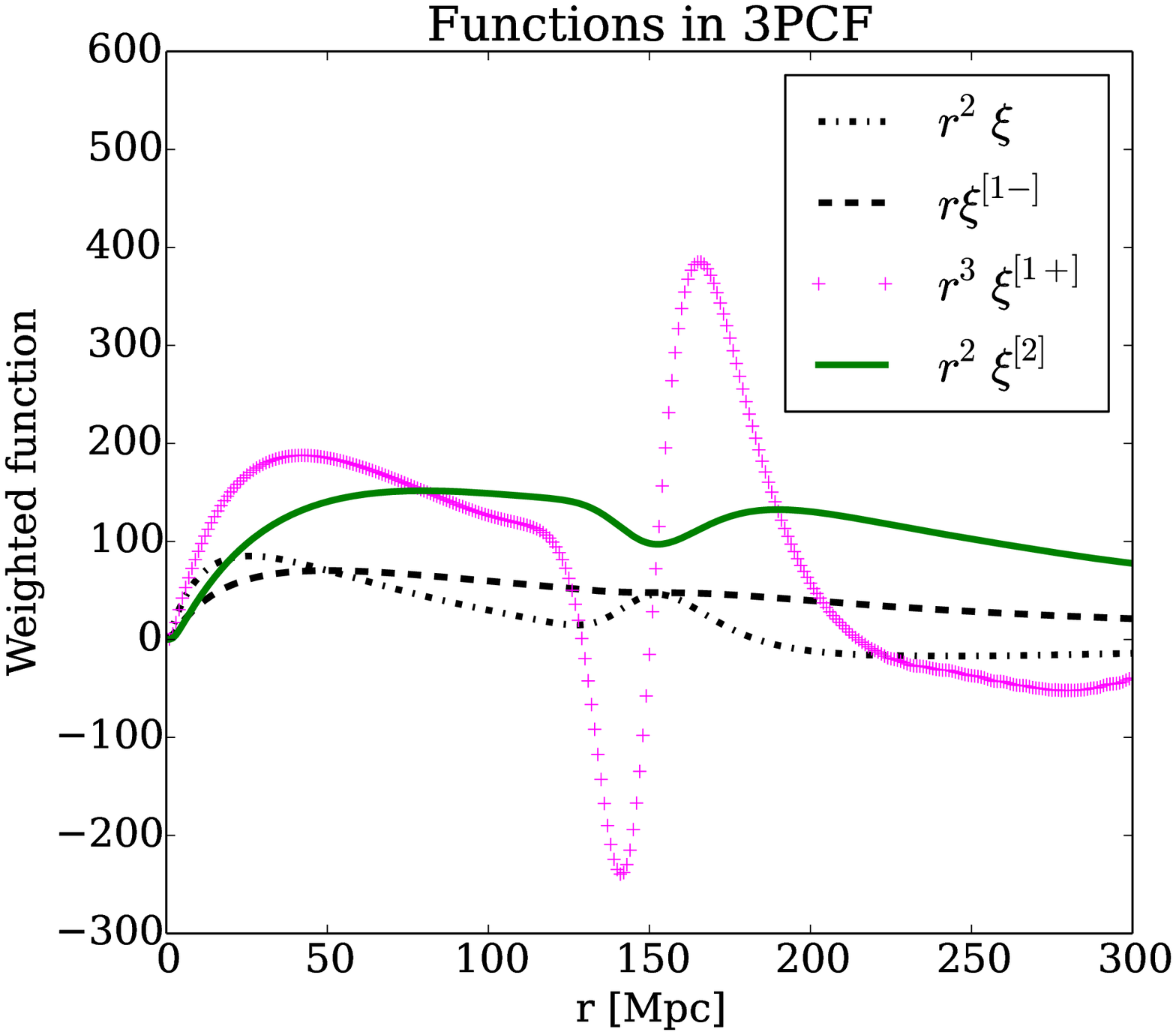}\hspace*{2in}\\
\includegraphics[width=0.68\textwidth]{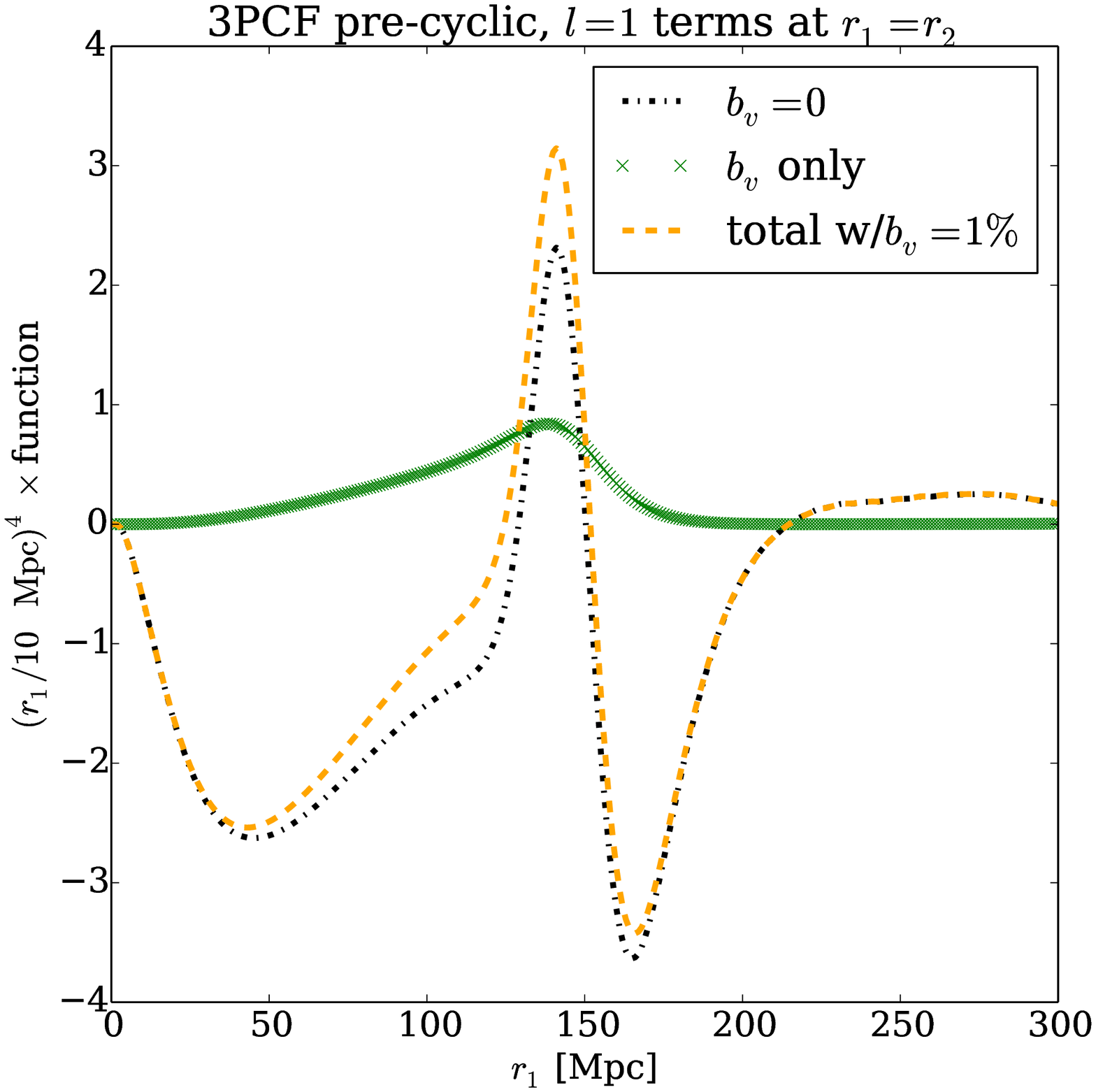}\hspace*{2in}
\end{center}
\caption{The top panel shows the functions appearing in the 3PCF pre-cyclic terms (equations (55) and (56)). Note the variety of structures around the acoustic scale. The bottom panel shows the products in the $l=1$
pre-cyclic term for isosceles triangles (i.e. $\zeta_{{\rm pc1}}\left(r_1,r_1\right)$).
Note that both $l=1$ terms have structure at the acoustic scale;
the no-velocity one inherits it from the BAO, while the velocity term
inherits it from the velocity Green's function (note the similarity
by comparing with Figure 2). We have used $b_{1}=1,b_{v}=0.01$.}
\end{figure}

Figure 8 shows the behavior of the coefficients that enter into the
Legendre polynomial expansion of the pre-cyclic term at $l=1$ (equation (53)), the
relevant multipole for the velocity. $\zeta_{{\rm pc1}}$ receives
an increment from the velocity for roughly isosceles triangles with
two sides of length $\sim r_{{\rm s}}.$ This can also be seen in the bottom panel fo Figure 7, which is a trace along the diagonals of the three panels in Figure 8.  The relative velocity effect
also produces a very modest decrement for triangles with one side
of length $\sim r_{{\rm s}}$ and one side of length $\sim2r_{{\rm s}}$.
The part of $\zeta_{{\rm pc}1}$ due to the usual (no relative velocity)
terms also has acoustic structure, with an increment for triangles
with one or more side of length $\sim r_{{\rm s}}$. Adding the velocity
(for $b_{v}>0$) and non-velocity contibutions to $\zeta_{{\rm pc1}}$
together produces a sharp increment for isosceles triangles with two
side lengths $\sim r_{{\rm s}}$, while the decrement from the relative
velocity for triangles with one side $\sim2r_{{\rm s}}$ is so modest
as to be washed out by the no-velocity contribution.

\begin{figure}
\includegraphics[scale=0.42]{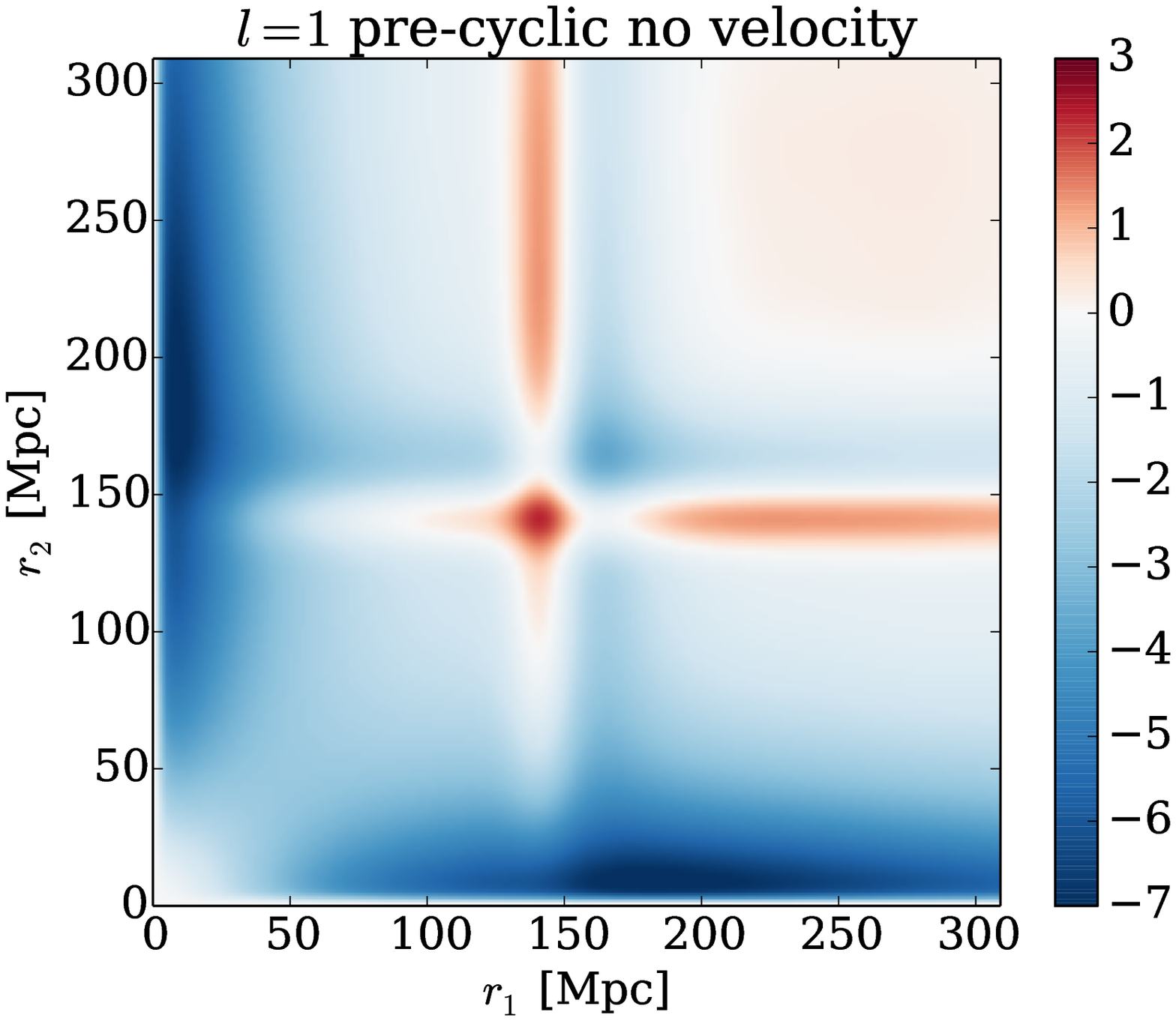}
\includegraphics[scale=0.42]{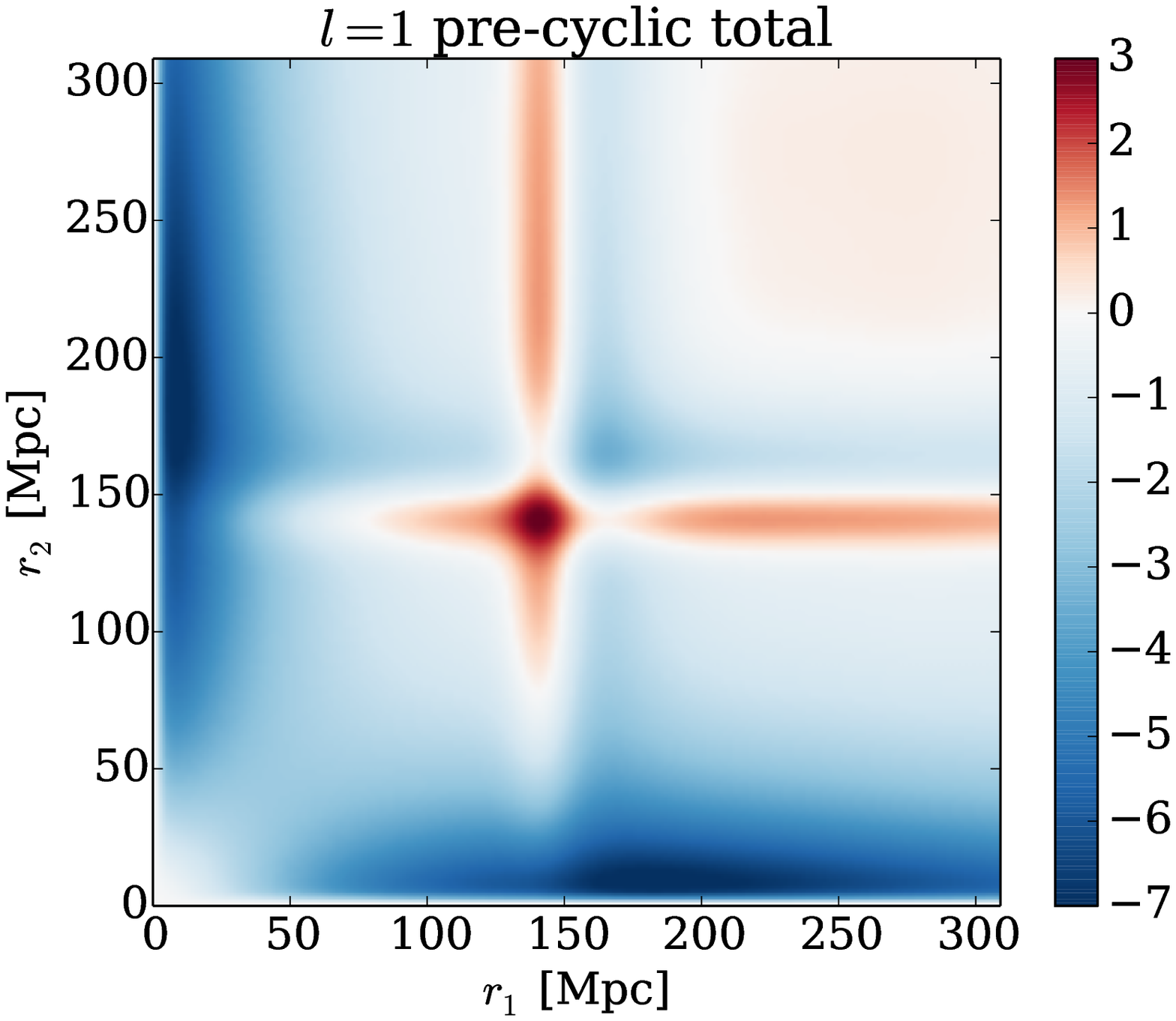}
\includegraphics[scale=0.42]{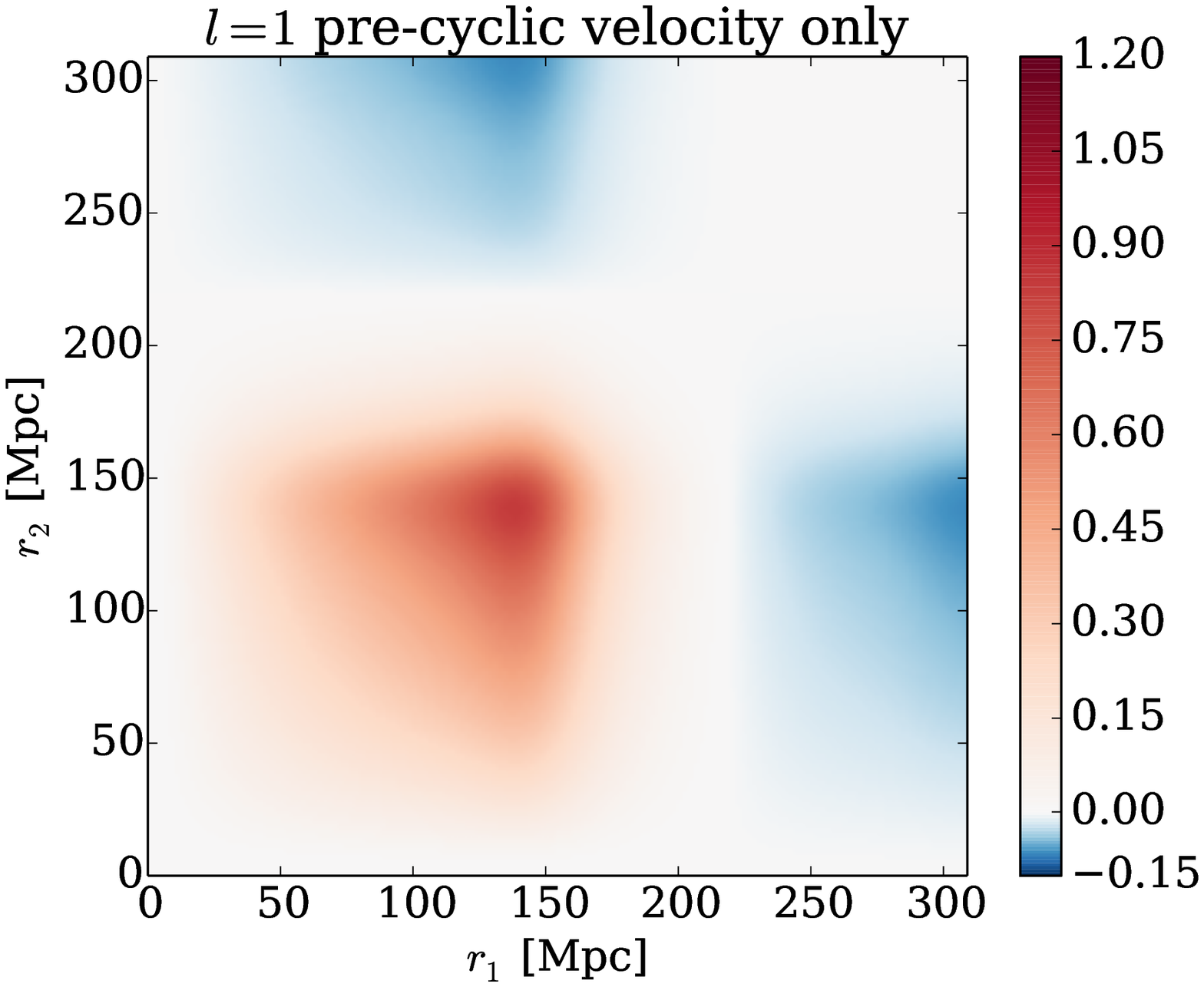}

\caption{The top panel shows the $P_{1}$ coefficient with $b_{v}=0$ (equation (53)). The middle panel shows the total $P_{1}$ coefficient
with velocity term included. The bottom panel shows the $P_{1}$ coefficient due to $b_{v}$
alone. Note that the relative velocity subtly enhances the number of triangles with two sides $\sim r_{\rm s}$ by carefully comparing the top two panels; this is made clear in the bottom panel.  We have used $b_{1}=1,b_{2}=0.1,b_{v}=0.01$ and weighted by $r_1^2r_2^2/10^4\;{\rm Mpc}^4$.}
\end{figure}

\section{Isolating $\lowercase{b}_{\lowercase{v}}$ from the 3PCF: cyclic summing and compression}

In \S6, we discussed the effect of the relative velocity on the 3PCF
with the simplification that we had chosen to evaluate $v_{s}^{2},\;\delta^2,$ and $\delta^{(2)}$
at the origin. This corresponds to our knowing which galaxy is contributing
these terms to the 3PCF; in practice, we do not know this.
Therefore, to give each of the three galaxies in a given triplet a
chance to contribute these terms, we must cyclically sum the
pre-cyclic 3PCF equation (50) around the triangle specified by $r_{1,}r_{2},$
and $\theta_{12}.$ We verify our prescription for this sum by calculating
the reduced 3PCF in a power-law $\xi\propto r^{-2}$ case and comparing
with Bernardeau et al. (2002)'s result (their equation (159) and Figure 10). After cyclically summing,
we re-project onto the basis of Legendre polynomials to find the radial coefficients for a multipole expansion of our 3PCF (the analog of equation (51)).  These are
\begin{align}
\zeta_{l}\left(r_{1},r_{2}\right)&=\frac{2l+1}{2}\int_{-1}^{1}d\mu_{12}\bigg[\zeta_{{\rm pc}}\left(r_{1},r_{2},\mu_{12}\right)\\
&+\zeta_{{\rm pc}}\left(r_{2},r_{3},\mu_{23}\right)+\zeta_{{\rm pc}}\left(r_{3},r_{1},\mu_{31}\right)\bigg]P_l(\mu_{12}),\nonumber
\end{align}
with $\mu_{12}\equiv\cos\theta_{12}$. Note that $r_{3,}\;\mu_{23}$ and $\mu_{31}$ are all functions of
$r_{1,}\; r_{2},$ and $\mu_{12}$, easily found using the law of
cosines. The factor of $\left(2l+1\right)/2$ is necessary because
$\int_{-1}^{1}P_{m}\left(\mu\right)P_{n}\left(\mu\right)d\mu=2/\left(2n+1\right)\delta_{mn}$;
that is, the Legendre polynomials are an orthogonal but not orthonormal
basis. Where in the pre-cyclic terms, we only found terms up to $l=2$, cyclically summing introduces higher orders. Indeed,
generically the cyclic sum projects onto an arbitrary number of Legendre
polynomials. We present the first few modes, split by bias coefficient,
in Figure 9.

In Figure 9 we must apply a more complicated weighting than our
usual $r^{2}$ weighting to the 3PCF. The 3PCF projections become
very large in magnitude for isosceles triangles. This is because when $\mu=1$, the Legendre polynomial being projected onto becomes very large.  This heavily weights squeezed triangles with zero opening angle.  When these triangles are also isosceles, their third side is zero, causing rapid increase in the functions of side length entering the pre-cyclic terms as these functions go roughly as $r^{-n},\;n>0$. These triangles are precisely those we must exclude, since for one side length $\lesssim 20\;\rm{Mpc}$ we expect perturbation theory to be invalid.  We have therefore suppressed
the diagonal by multiplying by a Gaussian weighting $\exp[-\left(12\;{\rm Mpc}/(r_{1}-r_{2})\right)^{2}]$.\footnote{Several choices of width for the Gaussian weighting were considered before choosing $12\;\rm{Mpc}$ as the best numerical factor above.}
We also weight by $r_{1}^{2}r_{2}^{2}/10^4\;{\rm Mpc}^4$ to make the fine structure
more apparent and to capture the expected contribution of each spherical
shell with volume $dV\propto r_{1}^{2}r_{2}^{2}$.

As we might expect, the velocity signature is strongest in $l=1$
but has echoes in $l=0$ and additional structure in $l>1.$ In particular,
the velocity bias produces in $\zeta_{1}$ an increment for triangles
with two sides $\lesssim r_{{\rm s}}$ and a decrement for triangles
with one side $\lesssim r_{{\rm s}}$ and one side $\gtrsim r_{s}$. This is consistent with what we might expect from the pre-cyclic term, which also has an increment and decrement for these respective configurations. Indeed, this can be roughly interpreted as a blurring of the structure present in the pre-cyclic $l=1$ velocity plot (compare Figure 8, bottom panel, with the $l=1$, $b_v$ panel in Figure 9).
Neither of the other bias coefficients contribute such an effect to
$\zeta_{1}$. The velocity bias produces in $\zeta_{2}$ a decrement
for triangles with one side $\lesssim r_{{\rm s}}$ and one side $\gtrsim r_{s}$.
Neither of the other bias coefficients contribute such an effect to
$\zeta_{2}$. The results for $\zeta_{3}$ and $\zeta_{4}$ are very
similar to those for $\zeta_{2}$.  This is because the higher-order Legendre polynomials are more sharply peaked at $\mu =\pm 1$, and so the triangles with $\mu=1$ (zero opening angle) and $\mu=-1$ are weighted heavily.\footnote{One can see this by noting that around $\mu=1$ the $P_l$s have series expansion $1-[l(l+1)/2](1-\mu)$ and so the drop in weight as $\mu$ decreases from 1 is more severe for higher $l's$.}  Thus as $l\to l+1$ for large $l$, the same small subset of triangles is dominating the projection integral (57), meaning the projections will be similar for $l$ and $l+1$.

The 3PCF decomposition we have presented thus far has two independent variables: the two triangle sides $r_1$ and $r_2$. It is worthwhile to consider whether this information can be compressed into a function of one independent variable.  By reducing the dimensionality of the problem, such a compression would ease handling of the covariance matrix associated with an eventual measurement, for example by accelerating the computations required in analyzing a large number of mock catalogs.  Further, a clever compression might allow us to avoid entirely the squeezed limit (isosceles triangles with small opening angle so that the third side approaches zero), in which perturbation theory is not expected to be valid because two of the galaxies are very close to each other.  Finally, a compression might allow us to focus on the set of triangles where the relative velocity is most pronounced: as Figure 9 and our previous discussion indicate, the relative velocity is localized to specific triangles in all $l$. In particular, on the scales expected to be better controlled observationally (i.e. those with $r_1$ and $r_2<200\;{\rm Mpc}$), the relative velocity is in a fairly condensed region of the $l=1$ multipole.  

With these desiderata in mind, we integrate the 3PCF's multipole moments (displayed with weighting in Figure 9) over one triangle side, but constrain this side to be within some fraction of the side that remains a free variable. We term this approach ``compression.''  If we integrate over $r_2\in [r_1/3,2r_1/3]$ and constrain $r_1>50\;{\rm Mpc}$, we can avoid any triangle side's nearing zero.  The minimum of $r_2$ will be $16.7\;{\rm Mpc}$, and by the Triangle Inequality the minimum of the third side, $r_3$, will be $33.3\;{\rm Mpc}$.  Thus all three triangle sides remain sufficiently large that perturbation theory should be valid. This compression scheme is shown in two different ways in Figure 10; the top panel portrays its effect in configuration space, while the bottom panel shows the region of each panel in Figure 9 integrated over.  Notice that the region of integraton captures much of the area where the relative velocity is important in $l=1$.  Several intervals for $r_2$ were considered before choosing $[r_1/3,2r_1/3]$; in an observational study the exact interval chosen might differ as optimality will somewhat depend on the signal-to-noise at different scales, an issue we do not treat in detail here.

Summarizing mathematically, to obtain the compressions at each $l$ split out by bias coefficient,
denoted $\bar{\zeta}_{lb_{1}},\bar{\zeta}_{lb_{2}},$ and $\bar{\zeta}_{lb_{v}}$, we compute
\begin{align}
\bar{\zeta}_{lx}=\frac{4\pi}{V_{{\rm shell}}}\int_{r_1/3}^{2r_{1}/3}r_{2}^{2}dr_{2}\zeta_{lx}(r_{1},r_{2}),
\end{align}
where $x$ ranges over $\left\{ b_{1},b_{2},b_{v}\right\}$ and $V_{{\rm shell}}=(7/27)(4\pi/3)r_1^3$ is the volume of the shell being integrated over.

Figure 11 shows that each $l$ we study can be used to detect the
relative velocity effect, with the strongest constraint coming from
$l=1,$ as might be expected given that this is the only mode where
the velocity contributes in the pre-cyclic 3PCF. It is encouraging
that $l=0$ through $l=2$ even show differences between the effect
and no-effect models at scales $\sim100\;{\rm Mpc},$ as this smaller
side length increases the number of triangles available in a given
survey volume relative to triangles with side length $r_{s},$ where
the effect is most pronounced in $l=1$. Interestingly, the effect
in all $l$'s studied at scales $r_{1}>r_{s}$ is substantial, though
survey volume limitations may not permit strong constraints to be
derived from such large scales. We have weighted the compressions
by $r_{1}$ to display the finer structure as well as to simulate
shot noise-limited measurements. Shot noise is inversely proportional to $\sqrt{N}$, the number of galaxies in a given spherical shell, so it scales as $1/r$.  Meanwhile the signal is proportional to the number of galaxies, so $S/N\propto r$.

\section{Discussion and conclusions}

Previous work by Yoo et al. (2011) has argued that the relative velocity effect can shift
the scale at which the BAO peak appears in the galaxy correlation
function $\xi_{{\rm gg}}$. In this work, we have shown that the relative
velocity's shift to $\xi_{{\rm gg}}$ is generated because the relative
velocity itself is non-zero only within the sound horizon at decoupling
(Figure 2). More precisely, we have explicitly calculated that the
velocity corrections to the correlation function are all generated
by convolving relatively narrow functions with a kernel that shares
the radial structure of the velocity Green's function. Hence the spatial
structure, in particular the compact support, of the Green's function
is inherited by the velocity corrections to $\xi_{{\rm gg}}$. Thus,
these corrections can alter the correlation function only below roughly
the sound horizon.\footnote{$\xi_{vv}$ has support out to $\sim 2r_{\rm s}$, but since this term scales as $b_v^2$ it is less important than $\xi_{v1}\propto b_v$ and $\xi_{v2}\propto b_v$, which drop nearly to zero outside $r_{\rm s}$.}
Adding or subtracting from the correlation function only below this
scale can change the radius at which the BAO peak occurs; a negative
velocity bias pushes it outwards while a positive bias pulls it inwards. 

To correct this shift and ensure that the BAO remain an accurate cosmological
ruler, the relative velocity bias must be measured. Motivated by the
previous work in Fourier space of Yoo et al. (2011), we have presented
a configuration-space template for fitting the three-point function
$\zeta$ to isolate $b_{v}.$ We have shown that the full 3PCF has
robust radial signatures of the relative velocity effect that are
unique and cannot come from any other bias term at the order in perturbation
theory to which we work, in agreement with Yoo et al. (2011). Furthermore, we have offered a useful basis
for measuring the full 3PCF and then suggested a further scheme for
processing these results. This scheme should unambiguously expose
the relative velocity's signature while avoiding the regimes in which
perturbation theory is expected to be inadequate. It will also likely ease handling of the covariance matrix if large numbers of mock catalogs are to be used for computing error bars.

Previous attempts to constrain the relative velocity observationally
have focused on Fourier space.  We have already alluded to the advantages of configuration space in \S1, and we revisit these points here.  On the theory
side, our configuration space approach has exposed the relatively
simple spatial structure of the relative velocity. Fourier-space work
on the bispectrum does not render transparent which triangle configurations
are optimal for velocity bias constraints. Our configuration
space approach immediately shows that, on scales small enough to measure with
current surveys, the velocity signature is localized to a small region
of triangle side lengths and a single multipole ($l=1$). This localized
signature (see Figure 9) naturally suggests that an integral over
the desirable region would enhance the velocity signal-to-noise,
an intuition borne out by our compression scheme results (Figure 11),
which, it should be noted, are weighted to reflect shot noise. 

On the practical side, there are also considerable advantages to working
in configuration space. As we have already discussed in \S1, edge-correction is much simpler in configuration space. Further, Pan \& Szapudi's (2005) measurement of the monopole moment of the 3PCF shows it is possible in practice to extract information from a multipole decomposition of the 3PCF.  Looking forward, in forthcoming work we will present a fast algorithm for computing the multipole moments of the 3PCF while accounting for edge correction.  This work will also address in more detail the covariance matrix of the 3PCF.

Thus far, the bispectrum technique of Yoo et al. (2011) has not been
used to constrain the relative velocity. Therefore three-point statistics
remain an entirely unexploited means of gaining traction on the relative
velocity, a situation we hope the configuration space signatures of
this work will improve. Nonetheless, recent work
by Yoo \& Seljak (2013) has used measurements of the power spectrum
in the Constant Mass (CMASS) sample of the Sloan Digital Sky Survey (SDSS) DR9 (260,000 galaxies) to compute a root-mean square shift of $0.57\%$ in the BAO peak position, showing this shift can
potentially be of order the entire error budget for the BAO distance
measurement. It should be noted that Yoo \& Seljak's best-fit parameter values imply no relative velocity at all (the root-mean square shift is from
integrating over the probability distribution of the linear, non-linear,
and velocity biases). On the other hand, a velocity bias of as large
as $\sim2\%$ (in our units; they use different values of $b_{1}$
and $b_{2}$ from our fiducial case so $b_{v}$ must be rescaled for
comparison) is consistent with their measurement at one sigma.\footnote{They find $b_1=2.2,\;b_2=0.65,\;b_v=0.037$, so we rescale such that $b_1=1$. Their definition of $b_2$ differs from ours by a factor of $2$ and $b_2=0.65$ is having accounted for this. This has also been done where we quote $b_2=0.1$ in the best-fit case from their work.} 

A limiting factor in their analysis is the growth of the error bars
at smaller wavenumber (large scales; see their Figure 6); as our analysis shows (see our Figure
6), large scales are important for the relative velocity's addition
to the correlation function. In this context two points should be
made. First, even when restricted to measurements of the power spectrum
(or correlation function), controlling the error bars on large scales
should have significant rewards, meaning increasing the number of
galaxies used for the measurement is highly desirable. Thus use of
the full sample of $\sim1$ million galaxies in the most recent SDSS
data should offer compelling new constraints. Second, the distinctiveness
of the 3PCF signature, where obtaining the dipole $(l=1)$ moment
already begins to isolate the velocity signature, should render the large-scale
error bars less problematic even at fixed number of galaxies.

Finally, a separate issue our work clarifies is how to constrain the
non-linear bias. Our compressions show that the higher multipoles
are extremely insensitive to the non-linear bias, while for $l=0$
and $l=1$ it contributes much more strongly than the linear bias
for a given magnitude of both $(b_{1}=b_{2}=1)$ (see Figure 12). Given that the non-linear bias enters only at $l=0$ in our pre-cyclic calculation (equation (52)), we indeed expect its projection onto $l=0$ and close-by multipoles to be strongest. This suggests that measuring different multipoles of the 3PCF, compressed
as we outline, should offer a robust way to separate the linear bias
from the non-linear bias. As the Yoo \&
Seljak (2013) best-fit measurement from CMASS shows, the non-linear
bias may be $\sim0.2$ in our units (their Figure 6), in which case Figure 11 shows it would contribute $\gtrsim20\%$ of what the linear bias does at $50\;{\rm Mpc}$ scales
in $l=0,$ but negligibly in $l=2$ and up. We hope the compression scheme
presented here will, independent of its utility for relative velocity
measurements, provide a new method to extract the non-linear bias
from 3PCF measurements.

Robustly separating the non-linear bias from other effects may also prove helpful in correcting the BAO peak shift if one is found.  Yoo \& Seljak (2013) find that non-linear bias $b_2$ can also shift the BAO peak. Historically $b_2$ has been constrained using the 3PCF or bispectrum (Scoccimarro et al. 2001; Verde et al. 2002; Wang et al. 2004; Gazta{\~n}aga et al. 2005; Gazta{\~n}aga et al. 2009; McBride et al. 2011b; Marin 2011; Marin et al. 2013; Guo et al. 2014), and our compressions offer a particularly clear way to isolate $b_2$ from the linear bias $b_1$ and from the velocity bias $b_v$.  This separation should aid accurate measurements of both $b_2$ and $b_v$ and ensure the peak shift can be corrected.

With the percent-level constraints on the cosmic distance scale the
BAO method offers through surveys such as BOSS (Anderson et al. 2014)
and the concomitant limits on dark energy's equation of state $w=p/\rho$
(Aubourg et al. 2014), understanding any sources of bias
is essential. In future work, we will implement the strategy discussed
here to measure $b_{v}$ using data from SDSS-III and assess whether
the imprint of the relative velocity between baryons and dark matter
might be present at a level relevant to modern BAO surveys.


\begin{figure*}
\centering\includegraphics[width=1.1\textwidth]{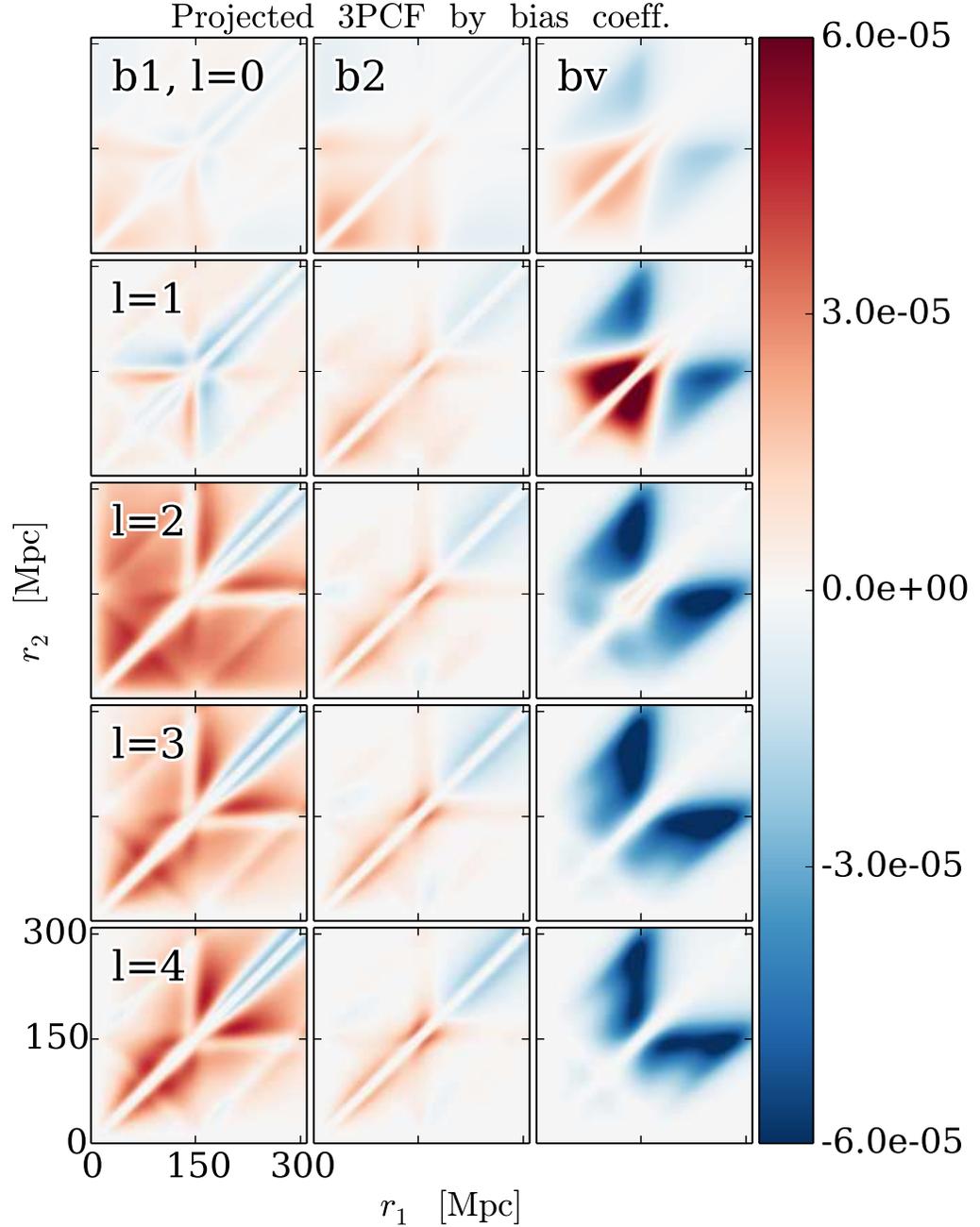}
\caption{Cyclically summed, weighted (see equation (57) and surrounding text) 3PCF split by bias coefficient and projected onto Legendre polynomials $P_{0}$ through $P_{4}$. Each row corresponds to one multipole, and the columns are as labeled at the top of the plot. The axes are $r_1$ (horizontal) and $r_2$ (vertical) in Mpc, and we have divided the linear bias ($b_1$) plots by $10$ so that all columns can be on the same colorbar. The weighting is $r_1^2r_2^2/10^4\;{\rm Mpc}^4\exp\left[-\left(12\;{\rm Mpc}/(r_{1}-r_{2})\right)^{2}\right]$; we have suppressed the diagonal because it is dominated by squeezed triangles for which perturbation theory is not valid.  We have used $b_{1}=1,b_{2}=0.1,b_{v}=0.01$. Note the distinctive
velocity signatures, especially in $l=1$: there is an increment for triangles with two sides $\lesssim r_{\rm s}$ and a decrement for triangles with one side $\lesssim r_{\rm s}$ and one side $\gtrsim r_{\rm s}$.}
\end{figure*}

\begin{figure}
\includegraphics[scale=0.5]{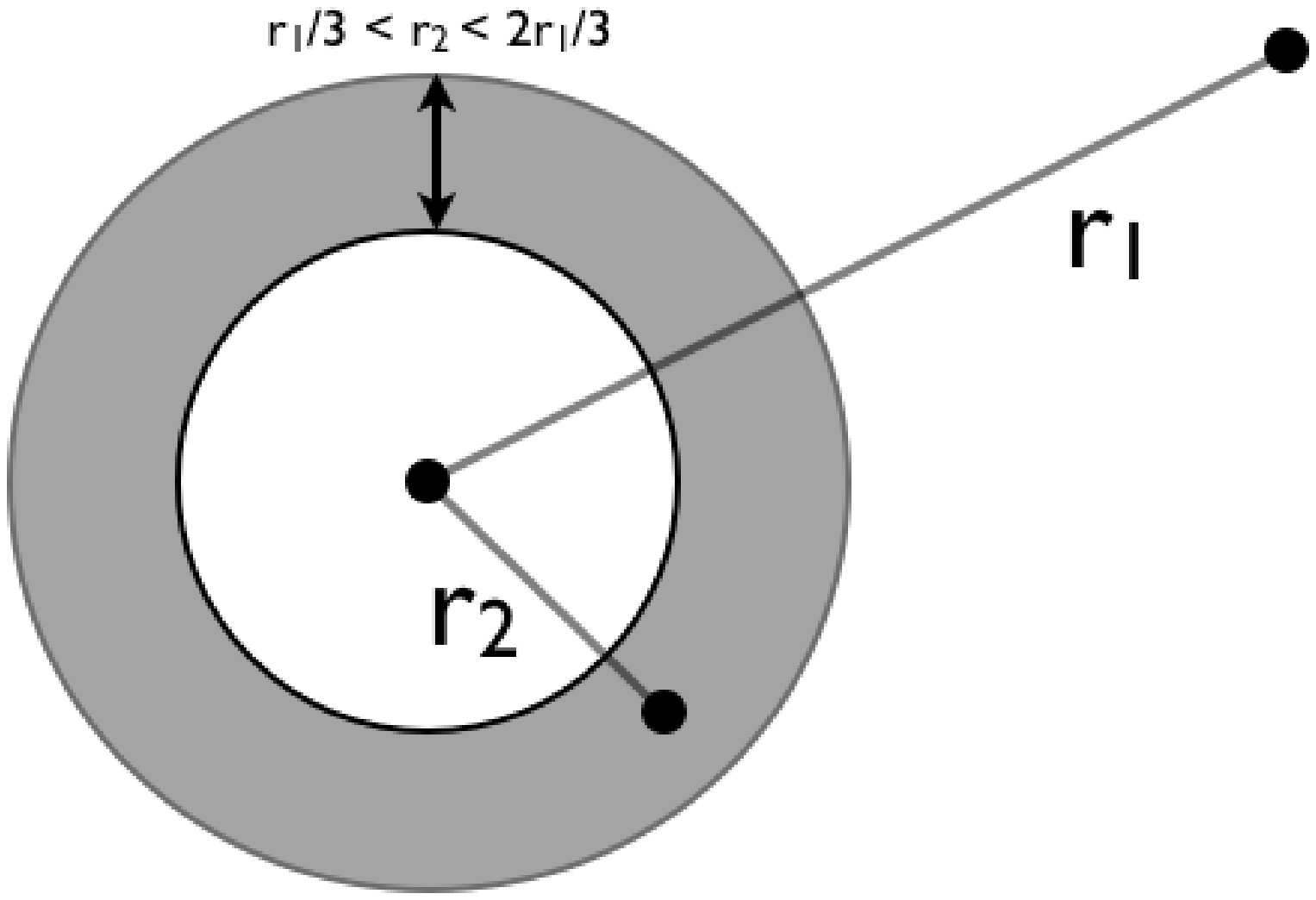}
\includegraphics[scale=0.4]{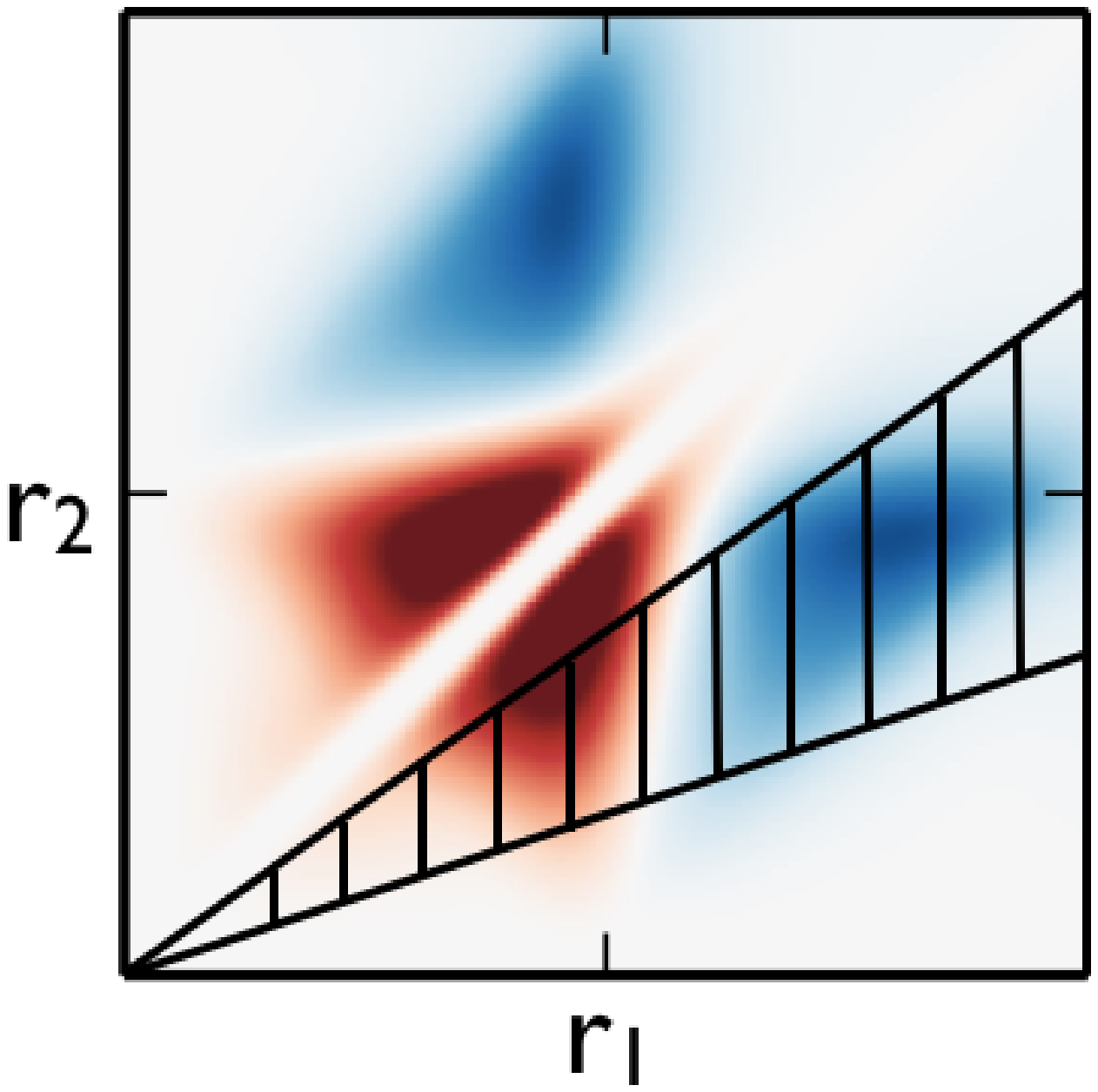}\caption{Compression scheme (see also equation (58)). Each dot represents a galaxy. We integrate the cyclically
summed 3PCF over all angles and over a spherical shell with $r_{2}$
from $\left(1/3\right)r_{1}$ to $\left(2/3\right)r_{1}$. This is
to capture the region of largest signal in Figure 9 while avoiding
the squeezed limit where one side of the triangle of galaxies becomes
so small as to invalidate linear perturbation theory. The region of
the $b_{v},$ $l=1$ plot being integrated over in Figure 9 is also shown above
as an example of what this scheme does to each sub-plot in Figure
9. Note that the diagonal is suppressed in Figure 9 but we do not apply this suppression when integrating for the compressions.}
\end{figure}

\begin{figure*}
\centering\includegraphics[width=1.\textwidth]{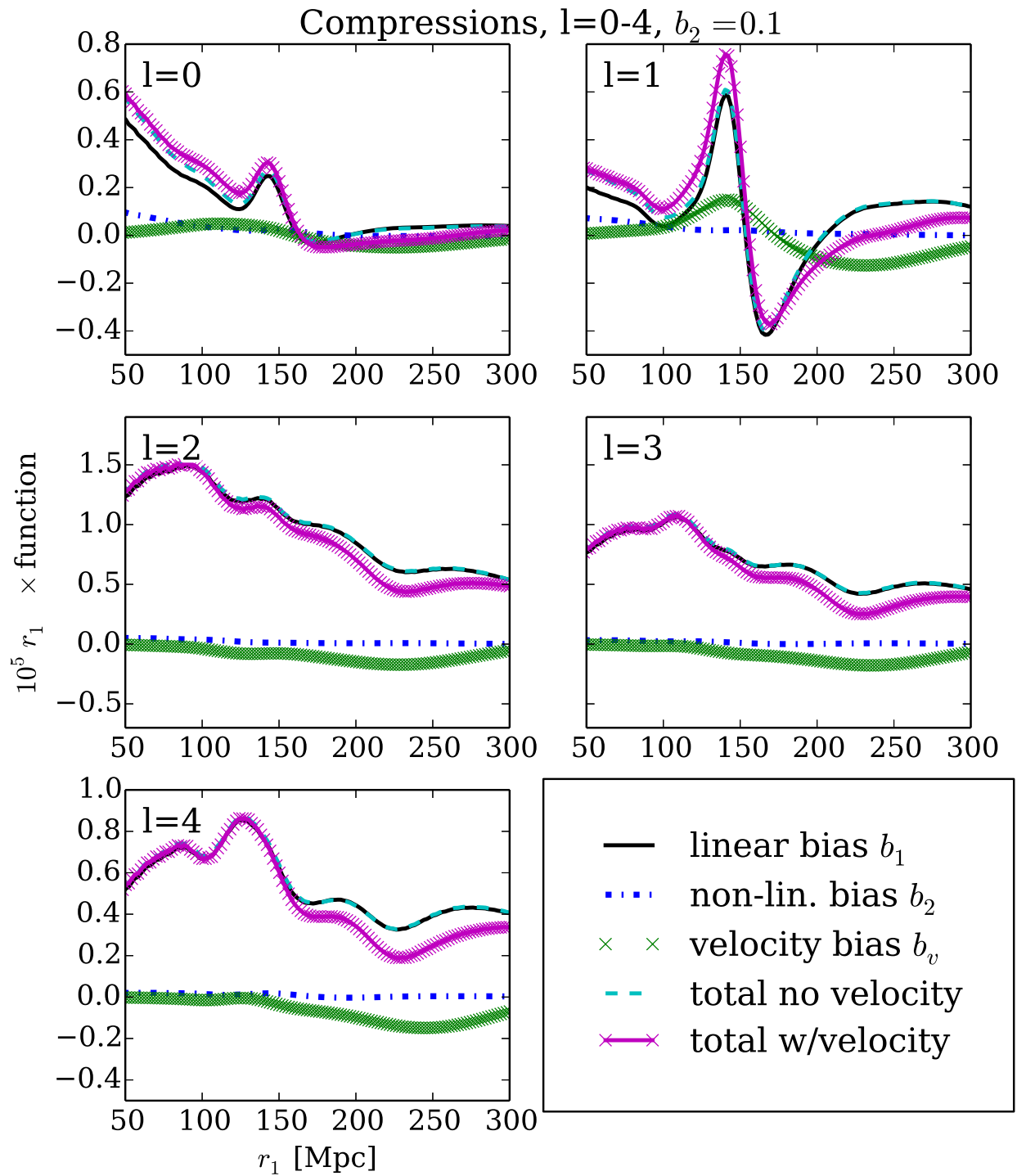}

\caption{Compressions by integrating along a wedge in Figure 9; see Figure 10 for visualization of this scheme. The no-velocity
model is in dashed cyan, while the model with velocity is in purple
X's. These show that the velocity produces distinctive signatures,
especially at $l=1$. Worth noting also is that $l=0$ and $l=1$ display
signal even at scales $r\sim100\;{\rm Mpc}<r_{s}$. These plots have
been weighted by $r_{1}$ to show the finer structure and model a
shot-noise limited measurement, and scaled up by $10^{5}$ to make
the vertical axis more compact. Notice that the non-linear bias (\textbf{$b_{2}$)
}does not contribute much to any of these plots, but especially so in the higher multipoles.}
\end{figure*}


\begin{figure*}
\centering\includegraphics[width=1.\textwidth]{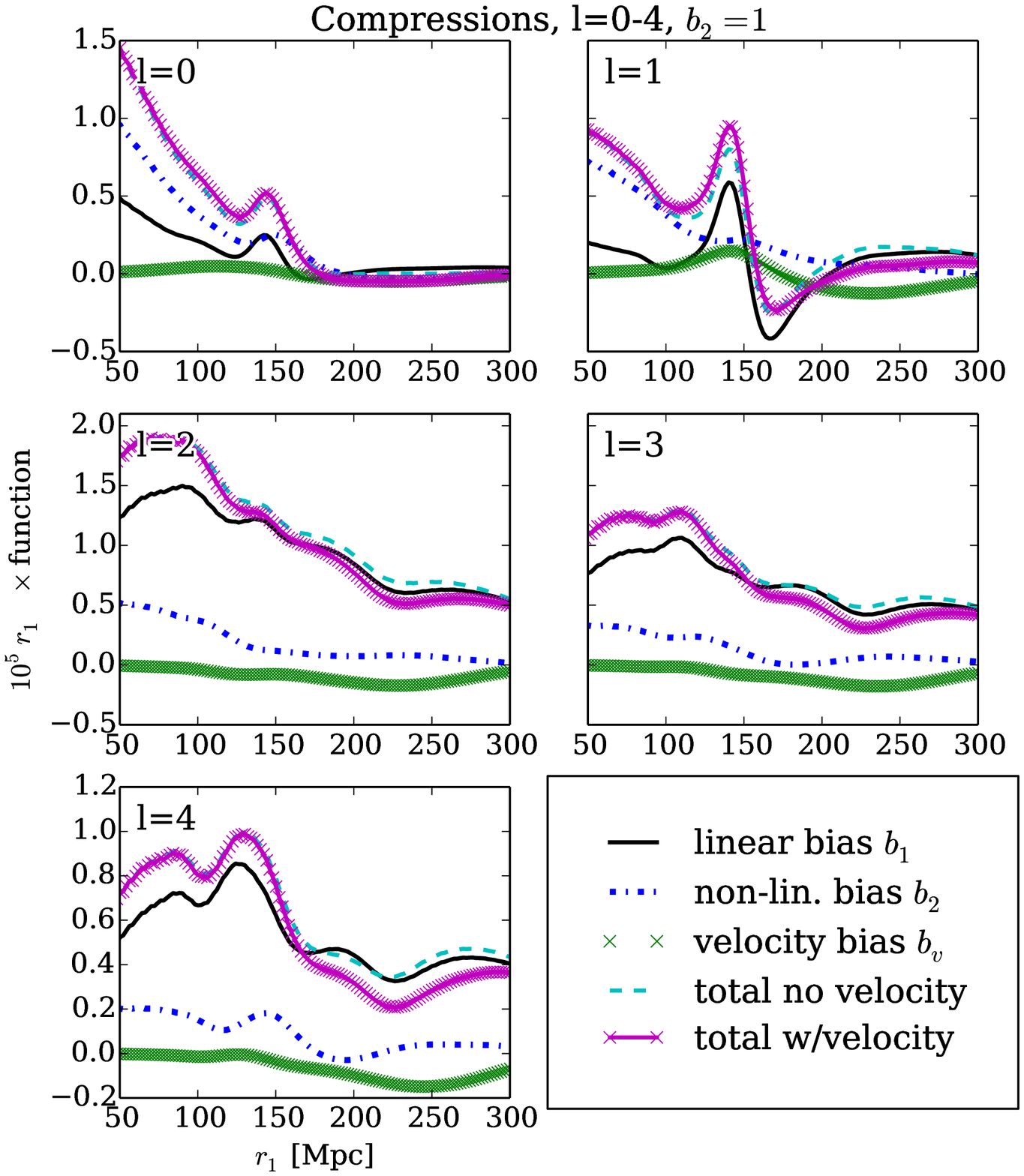}

\caption{To show what the contribution from $b_{2}$ would look like in an
extreme case, we present above results with $b_{1}=1,\; b_{2}=1,$ and
$b_{v}=0.01$. Even in this case, the relative velocity still has a distinct signature, especially in
$l=1$ if we focus on the $l$ for which the relative velocity has a signature on the smallest scales. Taking
the relative velocity aside, also note that here again we see that
the higher multipoles are much more sensitive to linear bias than
non-linear bias, providing a new means to discriminate between the
two and gain a robust measurement of each.}
\end{figure*}

\section*{Acknowledgements}

ZS thanks Neta Bahcall, Jo Bovy, Pierre Christian, Cora Dvorkin, Anastasia Fialkov, Doug Finkbeiner, Margaret Geller, JR Gott III, Laura Kulowski,
Avi Loeb, Chung-Pei Ma, Robert Marsland, Cameron McBride, Philip Mocz, Jim Moran, Stephen
Portillo, Matthew Reece, Mohammadtaher Safarzadeh, David Spergel, and Harvey Tananbaum for useful discussions
during this work.  ZS especially thanks Harry Desmond, Meredith MacGregor, Smadar Naoz, Yuan-Sen Ting, and Anjali Tripathi for comments on the manuscript. We also thank the anonymous referee for comments that considerably improved the scientific content and presentation of this work. This material
is based upon work supported by the National Science Foundation Graduate
Research Fellowship under Grant No. DGE-1144152.

\section*{References}

$\;\;\;\;\;$Adams JC, 1878, Proceedings of the Royal Society of London,
27, 63-71.

Albrecht A et al., 2006, Report of the Dark Energy Task Force, preprint
(astro-ph/0609591).

Anderson L et al., 2014, MNRAS, 441, 24.

Aubourg {\'E} et al., 2014, preprint (arXiv:1411.1074).

Barkana R, 2013, Pub. of the Astron. Soc. of Australia, 30.

Bashinksy S \& Bertschinger E, 2001, PRL 87, 8. 

Bashinsky S \& Bertschinger E, 2002, PRD 65, 12.

Bernardeau F, Colombi S, Gazta{\~n}aga E \& Scoccimarro R, 2002, Physics Reports, 367, 1-3, 1-248.

Bond JR \& Efstathiou G, 1984, ApJ 285, L45.

Bond JR \& Efstathiou G, 1987, MNRAS, 226, 655.

Bovy J \& Dvorkin C, 2013, ApJ 768, 1.

Chiba T, 2009, Phys. Rev. D, 79, 083517.

Chiba T, Dutta S \& Scherrer R, 2009, Phys. Rev. D, 80, 043517.

Chiba T, De Felice A \& Tsujikawa S, 2013, Phys. Rev. D, 87, 083505.

Copeland E, Sami M \& Tsujikawa S, 2006, Int. J. Mod. Phys. D, 15, 1753.

Dalal N, Pen U-L \& Seljak U, 2010, JCAP, 11, 7.

De Boni C, Dolag K, Ettori S, Moscardini L, Pettorino V \& Baccigalupi
C., 2011, MNRAS, 415, 2758.

Dutta S \& Scherrer R, 2008, Phys. Rev. D, 78, 123525.

Eisenstein DJ \& Hu W, 1997, ApJ 511, 5.

Eisenstein DJ \& Hu W, 1998, ApJ 496:605.

Eisenstein DJ et al., 2005, ApJ 633:560-574.

Eisenstein DJ, Seo H-J, Sirko E \& Spergel D, 2007b, ApJ 664:675-679.

Eisenstein DJ, Seo H-J \& White M, 2007a, ApJ 664:660-674.

Fialkov A, 2014, International Journal of Modern Physics D, 23, 8.

Fialkov A, Barkana R, Tseliakhovich D \& Hirata CM, 2012, MNRAS 424, 2.

Gazta{\~n}aga E, Cabr{\'e} A, Castander A, Crocce M \& Fosalba P, 2009, MNRAS 399, 2, 801-811.

Gazta{\~n}aga E, Norberg P, Baugh CM \& Croton DJ, 2005, MNRAS 364, 620-634.

Goroff MH, Grinstein B, Rey S-J \& Wise MB, 1986, ApJ 311, 6-14.

Greif TH, White SDM, Klessen RS \& Springel V, 2011, ApJ 736, 2.

Guo H, Li C, Jing YP \& B\"{o}rner G, 2014, ApJ 780, 139.

Hamilton AJS, 1998, ``The Evolving Universe'' ed. Hamilton D, Kluwer Academic, p. 185-275.

Holtzmann JA, 1989, ApJS, 71, 1.

Hu W \& Sugiyama N, 1996, ApJ 471:542-570.

Jain B \& Bertschinger E, 1994, ApJ 431: 495-505.

Kaiser N, 1987, MNRAS 227, 1-21.

Latif M, Niemeyer J \& Schleicher D, 2014, MNRAS 440, 4.

Li M, Li X \& Zhang X, 2011, Sci. China Phys. Mech. Astron., 53, 1631.

Kayo I et al., 2004, PASJ 56, 3.

Lewis A, 2000, ApJ 538:473-476.

Maio U, Koopmans LVE \& Ciardi B, 2011, MNRAS Letters, 412, 1, L40-44.

Marin F, 2011, ApJ 737:97.

Marin F et al., 2013, MNRAS 432, 4:2654-2668.

McBride C, Connolly AJ, Gardner JP, Scranton R, Newman J, Scoccimarro R, Zehavi I, \& Schneider DP, 2011a, ApJ 726:13.

McBride K, Connolly AJ, Gardner JP, Scranton R, Scoccimarro R, Berlind A, Marin F \& Schneider DP, 2011b, ApJ 739:85.

McQuinn M \& O'Leary RM, 2012, ApJ 760, 1.

Naoz S \& Barkana R, 2005, MNRAS 362, 3, 1047-1053.

Naoz S \& Narayan R, 2013, PRL 111, 5.

Naoz S, Yoshida N \& Barkana R, 2011, MNRAS 416, 1. 

Naoz S, Yoshida N \& Gnedin NY, 2012, ApJ 747, 2. 

Naoz S, Yoshida N \& Gnedin NY, 2013, ApJ 763, 1. 

O'Leary RM \& McQuinn M, 2012, ApJ 760, 1.

Pan J \& Szapudi I, 2005, MNRAS 362, 4, 1363-1370.

Peebles PJE \& Yu JT, 1970, ApJ 162, 815.

Richardson M, Scannapieco E \& Thacker R, 2013, ApJ 771:81.

Roukema BF, Buchert T, Ostrowski JJ \& France MJ, 2015, MNRAS, 448, 2, 1660-1673.

Scoccimarro R, Feldman HA, Fry JN \& Frieman JA, 2001, ApJ 546, 2, 652-664.

Seo HJ, Siegel ER, Eisenstein DJ, White M, 2008, ApJ 636:13-24.

Sherwin BD and Zaldarriaga M, 2012, Phys. Rev. D 85, 103523.

Silk J, 1968, ApJ 151:459-471.

Gott JR \& Slepian Z, 2011, MNRAS 416, 2, 907-916.

Slepian Z, Gott JR \& Zinn J, 2014, MNRAS 438, 3, 1948-1970.

Smith RE, Peacock JA, Jenkins A, White SDM, Frenk CS, Pearce FR, Thomas
PA, Efstathiou G, Couchman HMP, 2003, MNRAS 341, 1311.

Stacy A, Bromm V \& Loeb A, 2011, ApJ 730, 1.

Sunyaev RA \& Zel\textquoteright{}dovich Ya.B, 1970, Ap\&SS 7, 3.

Szapudi I \& Szalay A, 1998, ApJ 494, 1, L41-L44.

Tanaka T, Miao L \& Haiman Z, 2013, MNRAS 435, 4.

Tanaka TL \& Li M, 2014, MNRAS 439, 1, 1092-1100.

Szapudi I, 2004, ApJ 605:L89-92.

Tseliakhovich D \& Hirata CM, 2010, PRD 82, 083520.

Tseliakhovich D, Barkana R \& Hirata CM, 2011, MNRAS 418, 906.

Yoo J \& Seljak U, 2013, PRD 88:103520.

Yoo J, Dalal N \& Seljak U, 2011, JCAP 1107:018.

Verde L et al., 2002, MNRAS 335:432.

Visbal E, Barkana R, Fialkov A, Tseliakhovich D \& Hirata C, 2012, Nature, 487, 7405, 70-73.

Wang Y, Yang X, Mo HJ, van den Bosch FC \& Chu Y, 2004, MNRAS 353, 287-300.

Weinberg DH, Mortonson MJ, Eisenstein DJ, Hirata C, Riess AG \& Rozo
E, 2013, Physics Reports 530, 2, 87-255.

Wyithe JSB \& Djikstra M, 2011, MNRAS 415:3929-3950.

\section*{Appendix}

We begin by proving two theorems on the Fourier transform of a function
that can be represented as a Legendre polynomial of $\hat{r}_{1}\cdot\hat{r}_{2}$
times a function of $r_{1}$ and $r_{2}$, i.e. of the form
\begin{align}
K^{[l]}(\vec{r}_{1},\vec{r}_{2})&=K_{{\rm r}}^{[l]}\left(r_{1},r_{2}\right)P_{l}\left(\hat{r}_{1}\cdot\hat{r}_{2}\right)\\
&=K_{{\rm r1}}^{[l]}\left(r_{1}\right)K_{{\rm r2}}^{[l]}\left(r_{2}\right)P_l\left(\hat{r}_1\cdot\hat{r}_2\right),\nonumber
\end{align}
where in the last equality we additionally assume that the radial
piece $K_{{\rm r}}^{[l]}(r_{1},r_{2})$ can be split into a product of two functions with the same form.

We first show that the Fourier transform of a function of this form
will have the same angular dependence as the original function. We
define $\tilde{K}^{[l]}(\vec{k}_{1},\vec{k}_{2})={\rm FT}\left\{ K^{[l]}(\vec{r}_{1},\vec{r}_{2})\right\} (\vec{k}_{1},\vec{k}_{2}),$
where ${\rm FT}$ denotes a 6-D Fourier transform. We prove that
\begin{align}
\tilde{K}^{[l]}(\vec{k}_{1},\vec{k}_{2}) &=\left(-1\right)^{l}\mathcal{H}_{l}\left\{ K_{{\rm r}}^{[l]}\left(r_{1},r_{2}\right)\right\} \left(k_{1},k_{2}\right)\\
&\times P_{l}\left(\hat{k}_{1}\cdot\hat{k}_{2}\right).\nonumber
\end{align}
$\mathcal{H}_{l}$ is a 2-D transform defined as
\begin{align}
& \mathcal{H}_{l}\left\{ f\left(r_{1}\right)g\left(r_{2}\right)\right\}(k_1,k_2) =h_{l}\left\{ f(r_{1})\right\}(k_1) h_{l}\left\{ g(r_{2})\right\}(k_2);\nonumber\\ & h_{l}\left\{ f\left(r\right)\right\} \left(k\right)=4\pi\int dr r^{2}j_{l}\left(kr\right)f\left(r\right).
\end{align}
$\mathcal{H}_{l}^{-1}$ is defined analogously in terms of $h_{l}^{-1}$,
given by\footnote{Note that $h_{0}^{-1}\left\{ P\left(k\right)\right\} =\xi\left(r\right)$, and that the inverse's definition can be easily verified using equation (61) and the orthogonality relation for spherical Bessel functions.}
\begin{equation}
h_{l}^{-1}\left\{ \tilde{f}\left(k\right)\right\} \left(r\right)=\int \frac{k^{2}dk}{2\pi^2}j_{l}\left(kr\right)\tilde{f}\left(k\right).
\end{equation}
With this notation in place, we now prove equation (60). We will need
the following two identities. First, the spherical harmonic addition
theorem is
\begin{equation}
P_{l}\left(\hat{r}_{1}\cdot\hat{r}_{2}\right)=\frac{4\pi}{2l+1}\sum_{m=-l}^{l}Y_{lm}\left(\hat{r}_{1}\right)Y_{lm}^{*}\left(\hat{r}_{2}\right)
\end{equation}
where the $Y_{lm}$ are spherical harmonics and star means conjugate.
Second, the expansion of the plane wave in spherical harmonics is
\begin{equation}
e^{i\vec{k}\cdot\vec{r}}=4\pi\sum_{l=0}^{\infty}\sum_{m=-l}^{l}i^{l}j_{l}\left(kr\right)Y_{lm}\left(\hat{r}\right)Y_{lm}^{*}(\hat{k}).
\end{equation}
The 6-D Fourier transform of $K^{[l]}(\vec{r}_{1},\vec{r}_{2})$ is
\begin{equation}
\tilde{K}^{[l]}(\vec{k}_{1},\vec{k}_{2})=\int d^{3}\vec{r}_{1}d^{3}\vec{r}_{2}e^{i\vec{k}_{1}\cdot\vec{r}_{1}}e^{i\vec{k}_{2}\cdot\vec{r}_{2}}K_{{\rm r}}^{[l]}(r_{1},r_{2})P_{l}(\hat{r}_{1}\cdot\hat{r}_{2}).
\end{equation}
Using the spherical harmonic addition theorem (63) to replace $P_{l}(\hat{r}_{1}\cdot\hat{r}_{2})$
and applying the plane wave expansion (64) to replace the plane waves
we have
\begin{align}
&\tilde{K}^{[l]}(\vec{k}_{1},\vec{k}_{2})=\left(4\pi\right)^{3}\times\\
&\sum_{l_{1},m_{1},l_{2},m_{2}}\sum_{m=-l}^{l}\frac{i^{l_{1}+l_{2}}}{2l+1}Y_{l_{1}m_{1}}\left(\hat{k}_{1}\right)Y_{l_{2}m_{2}}^{*}\left(\hat{k}_{2}\right)\nonumber\\
&\times\int d^{3}\vec{r}_{1}d^{3}\vec{r}_{2}j_{l_{1}}\left(k_{1}r_{1}\right)j_{l_{2}}\left(k_{2}r_{2}\right)K_{{\rm r1}}^{[l]}\left(r_{1}\right)K_{{\rm r2}}^{[l]}\left(r_{2}\right)\nonumber\\
&\times Y_{lm}\left(\hat{r}_{1}\right)Y_{lm}^{*}\left(\hat{r}_{2}\right)Y_{l_{1}m_{1}}^{*}\left(\hat{r}_{1}\right)Y_{l_{2}m_{2}}\left(\hat{r}_{2}\right).\nonumber
\end{align}
Using the orthogonality of the spherical harmonics integrated over
solid angle we have
\begin{align}
&\tilde{K}^{[l]}(\vec{k}_{1},\vec{k}_{2})=\left(4\pi\right)^{3}\left(-1\right)^{l}\sum_{m=-l}^{l}\frac{1}{2l+1}Y_{lm}\left(\hat{k}_{1}\right)Y_{lm}^{*}\left(\hat{k}_{2}\right)\nonumber\\
&\times\int dr_{1}dr_{2}r_{1}^{2}r_{2}^{2}j_{l}\left(k_{1}r_{1}\right)j_{l}\left(k_{2}r_{2}\right)K_{{\rm r1}}^{[l]}\left(r_{1}\right)K_{{\rm r2}}^{[l]}\left(r_{2}\right).
\end{align}
We then have the desired equation (60) using in sequence equations
(63), (62), and (61). 

We now prove a useful result on the convolution of two functions of
the form given in equation (59). The second function is
\begin{equation}
Q^{[n]}(\vec{r}_{1},\vec{r}_{2})=Q_{{\rm r}}^{[n]}(r_{1},r_{2})P_{n}(\hat{r}_{1}\cdot\hat{r}_{2})
\end{equation}
with FT
\[
\tilde{Q}^{[n]}(\vec{k}_{1},\vec{k}_{2})=\left(-1\right)^{n}\mathcal{H}_{n}\left\{ Q_{{\rm r}}^{[n]}(r_{1},r_{2})\right\} (k_{1},k_{2})P_{n}(\hat{k}_{1}\cdot\hat{k}_{2})
\]
using equation (60). By the Convolution Theorem, $K^{[l]}\star Q^{[n]}$ is the
inverse FT of the product of the two functions' FTs, and using Adams'
(1878) identity for the product of 2 Legendre polynomials
\begin{equation}
P_{k}\left(x\right)P_{l}\left(x\right)=\sum_{m=|k-l|}^{k+l}\left(\begin{array}{ccc}
k & l & m\\
0 & 0 & 0
\end{array}\right)^{2}\left(2m+1\right)P_{m}\left(x\right)
\end{equation}
this product is
\begin{align}
&\tilde{K}^{[l]}(\vec{k}_{1},\vec{k}_{2})\tilde{Q}^{[n]}(\vec{k}_{1},\vec{k}_{2})=\\
&\left(-1\right)^{n+l}\mathcal{H}_{l}\left\{ K_{{\rm r}}^{[l]}(r_{1},r_{2})\right\} \mathcal{H}_{n}\left\{ Q_{{\rm r}}^{[n]}(r_{1},r_{2})\right\}\nonumber\\
&\times\sum_{m=|l-n|}^{l+n}\left(\begin{array}{ccc}
l & n & m\\
0 & 0 & 0
\end{array}\right)^{2}\left(2m+1\right)P_{m}(\hat{k}_{1}\cdot\hat{k}_{2}),\nonumber
\end{align}
where the $2\times3$ matrix is a Wigner 3j-symbol. Using the same
approach as for equation (60) to find the inverse FT we have
\begin{align}
&K^{[l]}(\vec{r}_{1},\vec{r}_{2})\star Q^{[n]}(\vec{r}_{1},\vec{r}_{2})=\\
&\sum_{m=|l-n|}^{l+n}\left(-1\right)^{n+l-m}\mathcal{H}_{m}^{-1}\left\{ \mathcal{H}_{l}\left\{ K_{{\rm r}}^{[l]}(r_{1},r_{2})\right\} \mathcal{H}_{n}\left\{ Q_{{\rm r}}^{[n]}(r_{1},r_{2})\right\} \right\}\nonumber\\
&\times\left(\begin{array}{ccc}
l & n & m\\
0 & 0 & 0
\end{array}\right)^{2}\left(2m+1\right)P_{m}\left(\hat{r}_{1}\cdot\hat{r}_{2}\right).\nonumber
\end{align}

\end{document}